\documentclass[twocolumn,10pt,prl,nofootinbib,preprintnumbers]{revtex4-1}
\usepackage[utf8]{inputenc}
\preprint{IPMU18-0094}
\preprint{DESY 18-076}

\def\mysections#1{{\bf #1.} }
\usepackage{hyperref}
\usepackage{cleveref}
\usepackage[dvips]{graphicx}
\usepackage{feynmp}
\usepackage{epsfig,amsmath,amssymb,verbatim,mathrsfs,array,layout,textcomp,amssymb,latexsym,slashed,graphicx,booktabs,color,mathtools,tikz}

\newcommand{\beq}{\begin{eqnarray}}
\newcommand{\eeq}{\end{eqnarray}}

\def\beqa{\begin{eqnarray}}
\def\eeqa{\end{eqnarray}}
\newcommand{\no}{\nonumber}
\newcommand{\bv}{\left(\begin{array}{c}}
\newcommand{\ev}{\end{array}\right)}
\newcommand{\bmtwo}{\left(\begin{array}{cc}}
\newcommand{\bmthree}{\left(\begin{array}{ccc}}
\newcommand{\emn}{\end{array}\right)}
\newcommand{\bmtwoc}{\left\{\begin{array}{cc}}
\newcommand{\bmthreec}{\left\{\begin{array}{ccc}}
\newcommand{\emnc}{\end{array}\right\}}
\newcommand{\ba}{\begin{array}}
\newcommand{\ea}{\end{array}}

\newcommand{\Tr}{{\text{Tr }}}

\def\lsim{\mathrel{\rlap{\lower4pt\hbox{\hskip1pt$\sim$}}
     \raise1pt\hbox{$<$}}}         
\def\gsim{\mathrel{\rlap{\lower4pt\hbox{\hskip1pt$\sim$}}
     \raise1pt\hbox{$>$}}}         

\begin{document}
\font\mini=cmr10 at 0.8pt

\title{
Twin SIMPs
}

\author{Yonit Hochberg${}^{1}$}\email{yonit.hochberg@mail.huji.ac.il}
\author{Eric Kuflik${}^{1}$}\email{eric.kuflik@mail.huji.ac.il}
\author{Hitoshi Murayama${}^{2,3,4,5}$}\email{hitoshi@berkeley.edu, hitoshi.murayama@ipmu.jp}
\affiliation{${}^1$Racah Institute of Physics, Hebrew University of Jerusalem, Jerusalem 91904, Israel}
\affiliation{${}^2$Ernest Orlando Lawrence Berkeley National Laboratory, University of California, Berkeley, CA 94720, USA}
\affiliation{${}^3$Department of Physics, University of California, Berkeley, CA 94720, USA}
\affiliation{${}^4$Kavli Institute for the Physics and Mathematics of the
  Universe (WPI), University of Tokyo,
  Kashiwa 277-8583, Japan}
\affiliation{${}^5$DESY, Notkestra\ss e 85, D-22607 Hamburg, Germany}

\preprint{DESY 18-076, IPMU18-0094}

\begin{abstract}

The hierarchy problem and the identity of dark matter are two of the central driving forces in particle physics. Twin Higgs models provide an elegant solution to the little hierarchy problem, while Strongly Interacting Massive Particles (SIMPs) provide an appealing dark matter candidate. Here we show that SIMPs can easily be embedded in the Twin Higgs setup, such that dark matter and the hierarchy problem can be addressed in a single framework.  This also provides a natural explanation to the proximity between the confinement scale of SIMP dark matter and the strong scale of QCD.
\end{abstract}

\maketitle

\section{Introduction}

The hierarchy problem between the electroweak scale and higher energy scales is one of the most pressing problems in particle physics. The discovery of the Higgs boson at the Large Hadron Collider (LHC) in 2012 along with the lack of experimental evidence for new weak-scale particles highlights this problem further. 
Indeed, recent years have sparked new directions for solving the hierarchy problem. 
For instance, Twin Higgs models address the hierarchy problem without introducing new particles charged under the Standard Model (SM)~\cite{Chacko:2005pe}. The presence of a mirror sector with its own gauge group of $SU(3)\times SU(2) \times U(1)$ is assumed, with a ${\mathbb Z}_2$ symmetry that relates the SM and mirror twin sectors acting to protect the Higgs mass from quadratic divergences at one loop. Small breaking of the ${\mathbb Z}_2$ symmetry is needed in order to obtain a phenomenologically viable Higgs sector. (For other variations of neutral naturalness, see {\it e.g.} Refs.~\cite{Chacko:2005vw,Batell:2015aha,Burdman:2006tz,Cai:2008au,Craig:2014aea,Craig:2014roa,Arkani-Hamed:2016rle}.) Thus twin Higgs models (and other theories of neutral naturalness) contain a QCD-like sector, similar to QCD of the SM but not identical to it.

The search for the identity of dark matter is also one of the greatest mysteries of modern physics. The lack of experimental observation of Weakly Interacting Massive Particles (WIMPs) has led in recent times to a surge of new ideas for dark matter with various mass scales and interactions. 
The Strongly Interacting Massive Particle (SIMP)~\cite{Hochberg:2014dra} is such a dark matter candidate, where the freeze-out of $3\to2$ self-annihilations set the relic abundance. The kinetic energy generated in the system by this process must be shed, which can be achieved via
thermalization between the dark and visible sectors.  
The SIMP setup then predicts dark matter of order a few hundreds of MeV, with strong self-interactions and very weak couplings to the visible sector.  Importantly, SIMPs are generic in QCD-like theories of dynamical chiral symmetry breaking, with the pseudo-Nambu-Goldstone bosons playing the role of dark matter~\cite{Hochberg:2014kqa}. The Wess-Zumino-Witten term~\cite{Wess:1971yu,Witten:1983tw,Witten:1983tx} generates the requisite $3\to2$ self-interactions~\cite{Hochberg:2014kqa}, and the equilibration between the dark sector and the SM can be obtained {\it e.g.} via a kinetically mixed hidden photon~\cite{Lee:2015gsa,Hochberg:2015vrg}, with many novel experimental signatures~\cite{Lee:2015gsa,Hochberg:2015vrg,Berlin:2018tvf}.

Since SIMPs are generic in QCD-like sectors, and twin Higgs theories contain a QCD-like sector, it is appealing to try and merge these two notions into one: namely, to embed SIMP dark matter into the QCD sector of twin Higgs. 
Then, SIMPs would be the pseudo-scalar mesons in the twin QCD sector.  In what follows, we show that this can easily be done, obtaining a natural theory with strongly self-interacting sub-GeV dark matter while addressing the hierarchy problem.  (For other works on dark matter in twin Higgs models, see {\it e.g.} Refs.~\cite{Garcia:2015loa,Craig:2015xla,
Garcia:2015toa,Farina:2015uea}.)

Moreover, the similarity of the confining scales between QCD and the SIMP dynamics, as is requisite for a SIMP, is naturally explained in such a framework: Since twin QCD effects enter the Higgs potential at two loops~\cite{Craig:2015pha}, the twin Higgs solution to the hierarchy problems requires the twin QCD coupling and scale---namely the SIMP sector in the framework presented here---to be similar to that of the QCD sector itself. 

\section{Concept}

The ingredients of our setup are as follows. The lightest particles in the twin sector are the twin mesons, which are identified with SIMP dark matter. All other twin particles annihilate or decay into these twin mesons or SM particles. The twin fermions are heavier than the light twin quarks that form the SIMP dark matter. Since all Yukawa couplings except for that of the top quark are irrelevant to the hierarchy problem due to their smallness, this spectrum does not spoil the twin Higgs mechanism. The twin photon and twin neutrinos (be they Dirac or Majorana) are likewise heavy, and can decay away. The typical cosmological problem of twin Higgs models, where too-large contributions to $N_{\rm eff}$ often arise, is thus naturally absent here.

Amongst the first two generations of twin quarks, we impose an exact global $SU(2)_f$ symmetry. The lightest twin mesons are a flavor triplet $\left(d'\bar{s}', s'\bar{d}', \frac{1}{\sqrt{2}}(s'\bar{s}'-d'\bar{d}')\right)$, which we call {\it pions},~$\pi$.  They are stable since they are the lightest particles with a conserved $SU(2)_f$ quantum number.  Here and below, we denote particles in the twin sector with a prime on the corresponding SM particles, except for the twin mesons, further defined below.

\begin{figure}[t!]
\begin{center}
\includegraphics[width =0.49\textwidth]{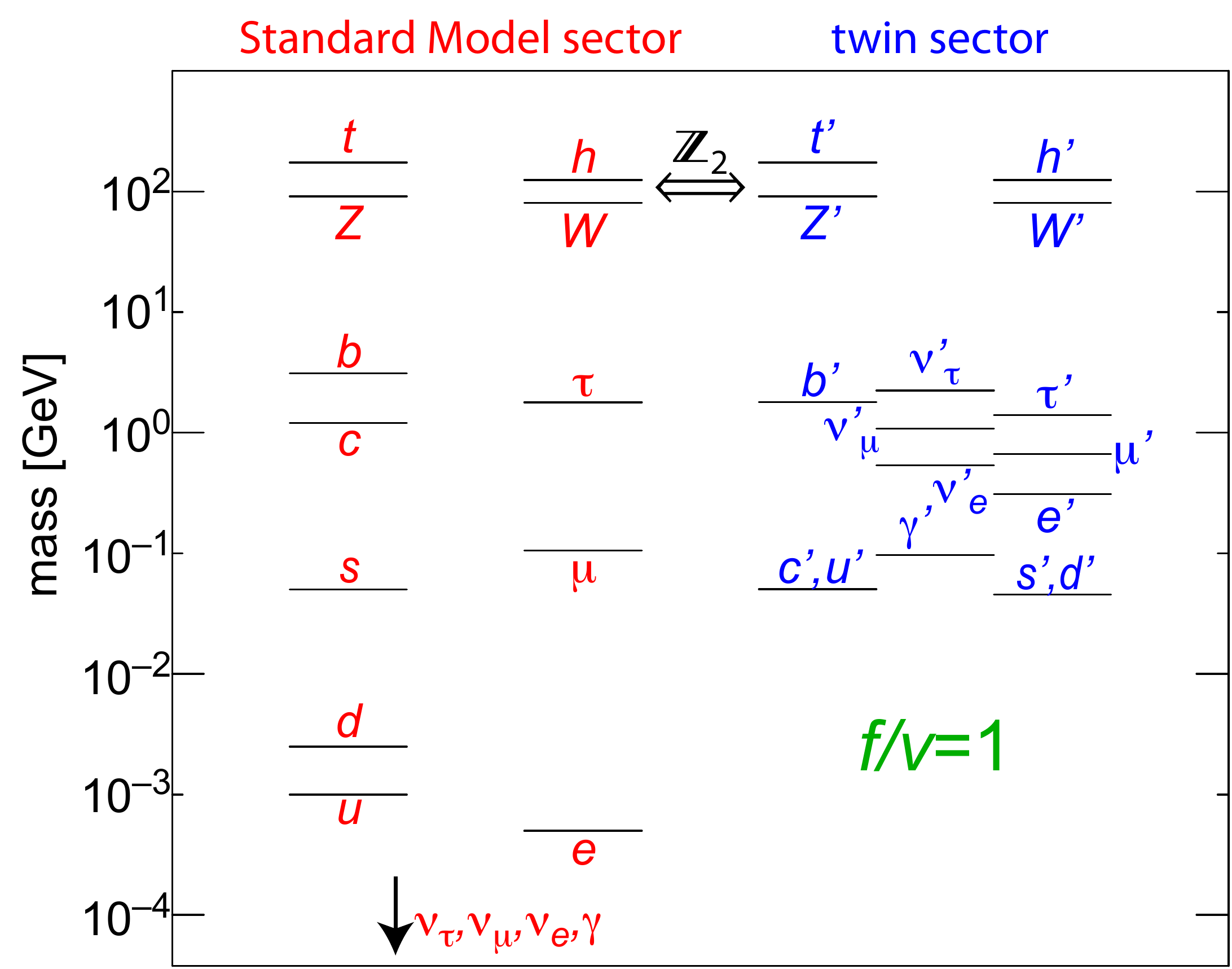}
\end{center}
\caption{\label{fig:spectrum}
A sample spectrum of twin particles.  Here we use $f/v=1$ to demonstrate the ${\mathbb Z}_2$ invariance between the visible and twin sectors for $t$, $h$, $Z$, $W$; lighter particles are subject to ${\mathbb Z}_2$-breaking effects without spoiling the solution to the hierarchy problem.  In practice, twin sector masses are raised by a factor of $f/v \gtrsim 3$.
}
\end{figure}

\section{Thermal History}

A simple example of a twin mass spectrum for our framework is shown in Fig.~\ref{fig:spectrum}.  The twin particles at the electroweak scale ---$W'$, $Z'$, $t'$, $h'$--- have similar masses to their visible sector counterparts due to the ${\mathbb Z}_2$ symmetry.  In practice, the ratio of vacuum expectation values between the twin and SM sectors is $f/v\gtrsim 3$ and the twin particles are heavier by the common factor.  In the early Universe, they decay away quickly.  The neutrinos also decay, $\nu_l' \rightarrow l' u' \bar{d}', l' c' \bar{s}'$.  The bottom quark and charged leptons annihilate away $b' \bar{b}' \rightarrow g' g', q'\bar{q}'$, $l^{\prime +} l^{\prime -} \rightarrow \gamma' \gamma',q'\bar{q}'$, with negligible abundances.  The heavy meson abundances are likewise negligible (see Ref.~\cite{Geller:2018biy} for a detailed analysis). The twin photon is also massive (as can be achieved via the St\"uckelberg mechanism for the $U(1)'_Y$ gauge boson).  At temperatures of order the GeV-scale, only four light twin quarks, the twin gluons, and possibly the massive twin photon are around.

The global $SU(2)_f$ invariance dictates $m_{u'} = m_{c'}$, $m_{d'} = m_{s'}$.  We arbitrarily take $m_{d',s'} < m_{u',c'} = m_{d',s'} (1+2\Delta) $, with a mass splitting $\Delta \lesssim 10$\%.  An approximate $SU(4)_f$ flavor symmetry for the twin QCD exists in addition to the twin $U(1)_{\rm EM}$, and is broken to $SU(2)_U \times SU(2)_D \times U(1)_{\rm EM}$ by $\Delta$.  The two $SU(2)$'s are broken to the diagonal subgroup $SU(2)_f$ by the twin weak interaction $SU(2)_L$, and the remaining global symmetry is $SU(2)_f \times U(1)_{\rm EM}$. 

\begin{figure}[t!]
\begin{center}
\includegraphics[width =0.4\textwidth]{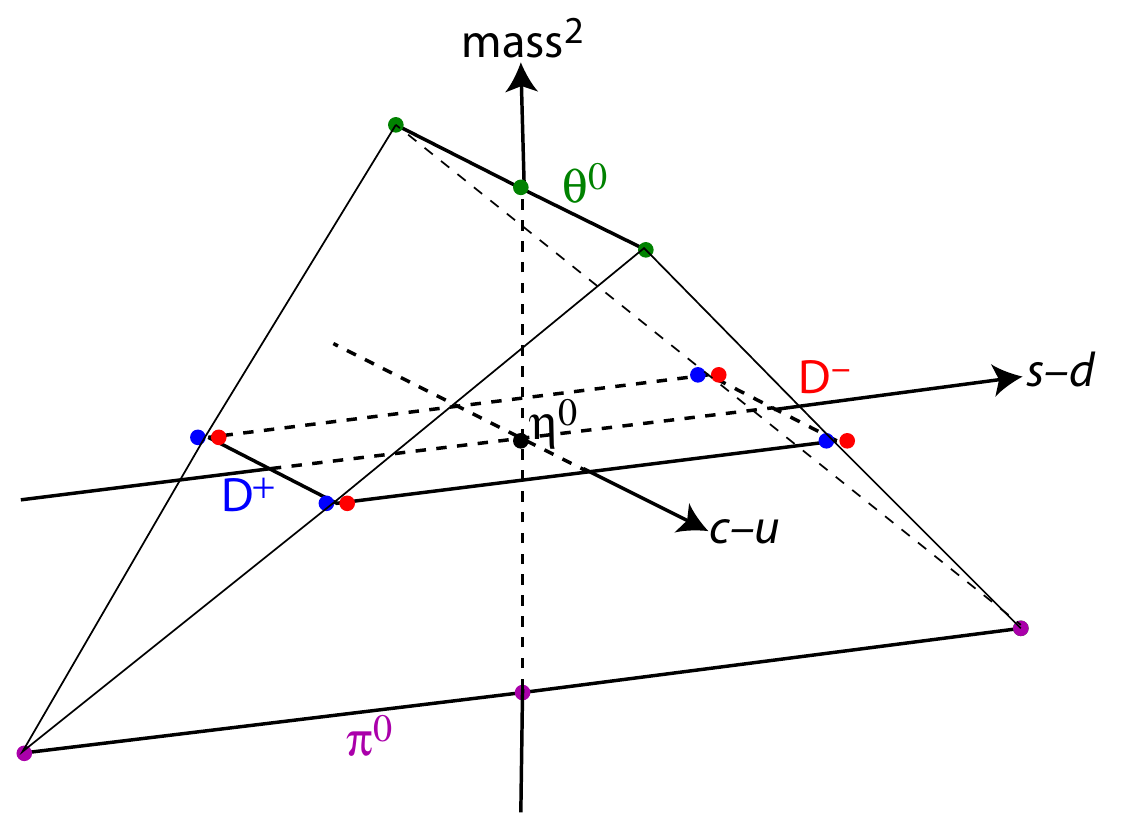}
\end{center}
\caption{\label{fig:mesons}
A visual representation of the meson spectrum.
}
\end{figure}

Twin QCD confines and produces a 15-plet of mesons~$M$ in the adjoint representation of the approximate $SU(4)_f$ symmetry.  Table~\ref{tab:M} shows the meson decomposition, as well as  the combination of quark masses that generates the masses-squared of the mesons.  The lightest meson states, which are the pions $\pi$,  are the SIMP dark matter. A visual representation of the meson spectrum is given in Fig.~\ref{fig:mesons}.

\begin{table}[t!]
\begin{tabular}{|c|c|c|c|}
\hline
meson $M$ & particle content & $m_M^2 \propto$ & $m_M$ \\ \hline
$\theta^0 ({\bf 3}, {\bf 1})$ & $u' \bar{c}', c' \bar{u}', \frac{1}{\sqrt{2}} (u'\bar{u}'-c'\bar{c'})$ & $2m_{u'}$ & $m_\pi(1+\Delta)$\\
$D^+ ({\bf 2}, {\bf 2})$ & $u' \bar{d}', c' \bar{d}', u'\bar{s}', c'\bar{s}'$ & $m_{u'} + m_{d'}$ & $m_\pi (1+\frac{\Delta}{2})$\\
$D^- ({\bf 2}, {\bf 2})$ & $d' \bar{u}', s' \bar{u}', d'\bar{c}', s'\bar{c}'$ & $m_{u'} + m_{d'}$ & $m_\pi (1+\frac{\Delta}{2})$\\
$\eta^0 ({\bf 1}, {\bf 1})$ & $\frac{1}{2} (d' \bar{d}' + s' \bar{s}' - u'\bar{u}' - c'\bar{c}')$ & $m_{u'} + m_{d'}$ & $m_\pi (1+\frac{\Delta}{2})$\\
$\pi^0 ({\bf 1}, {\bf 3})$ & $d' \bar{s}', s' \bar{d}', \frac{1}{\sqrt{2}} (d'\bar{d}'-s'\bar{s}')$ & $2m_{d'} $ & $m_\pi$\\ \hline
\end{tabular}
\caption{Decomposition of the meson $SU(4)_f$ 15-plet under $SU(2)_U \times SU(2)_D \times U(1)_{\rm EM}$.  The 3$^{\rm rd}$ column shows the linear combination of quark masses that determines the meson masses-squared. The 4$^{\rm th}$ column shows the mass splittings. From top to bottom, the meson masses go from heaviest to lightest, assuming $m_{d'} = m_{s'} < m_{u'} = m_{c'}=m_{d',s'}(1+2\Delta)$.}\label{tab:M}
\end{table}

We note that the global $SU(2)_f$ symmetry forbids Cabbibo--Kobayashi--Maskawa (CKM) mixing among twin quarks.  As a result, twin generation number is conserved in this setup.

The twin mesons undergo $3\rightarrow 2$ annihilations~\cite{Hochberg:2014dra,Hochberg:2014kqa} via the Wess--Zumino--Witten action of the $SU(4)_f$ chiral Lagrangian~\cite{Wess:1971yu,Witten:1983tw,Witten:1983tx}: 
\beq
{\cal L}_{3\to2}
  = 
  \frac{2}{5 \pi^2 f_\pi^5}\epsilon^{\mu\nu\rho\sigma} \Tr (\pi \partial_\mu
  \pi \partial_\nu \pi \partial_\rho \pi \partial_\sigma \pi )\, .
  \eeq
The meson mass splittings are given by $\sim(\frac{1}{2}-1)\Delta \lesssim 5-10$\% so that all 15 mesons can co-annihilate at the freeze-out temperature $T_f = m_\pi / x_f \approx m_\pi / 20$. 
The observed dark matter relic abundance is obtained for twin pion masses $m_\pi$ of order a few hundred MeV, in the strongly interacting regime of the theory, $m_\pi/f_\pi \sim 2\pi$~\cite{Hochberg:2014kqa}. Strong self-scattering cross sections, relevant for puzzles in structure formation, are thus expected as well. The above features persist even in the presence of small mass splittings amongst the mesons. For further details, see Ref.~\cite{Hochberg:2014kqa}.

During dark matter freezeout, kinetic mixing $\epsilon$ between the twin photon $\gamma'$ and the SM $\gamma$ (sourced by mixing with the twin and SM hypercharge gauge bosons),  maintains  thermal equilibrium between the two sectors via the scattering process of twin mesons $M$ off of electrons, $M e \rightarrow M e$.  The allowed parameter space is similar to that studied in detail in Ref.~\cite{Hochberg:2015vrg}: Twin photon masses between $\sim 2m_\pi$ and 100's of GeV are viable over a broad range of $\epsilon$ values, where kinetic equilibrium between the twin and SM sectors is maintained while the annihilations $M+M \rightarrow {\rm SM}$ through $\gamma'$ are suppressed compared to the $3\to2$ annihilations.  Multiple future experimental probes are set to test this parameter space (see further discussion below). Note that twin elastically decoupling relic (ELDER) DM~\cite{Kuflik:2015isi,Kuflik:2017iqs}  can also be realized in our framework, where the relic density of twin mesons is set by the elastic scattering off of the electrons.

In the twin sector, some of the heavier mesons are unstable against decays. Among the 15-plet of mesons, the $\pi$, $D^+$ and $D^-$ are stable because of their conserved quantum numbers: the $\pi$'s are the lightest particles with a non-trivial $SU(2)_f$ quantum number, while the $D^\pm$'s are the lightest particles charged under the twin QED. The $\theta$ and $\eta$ are, in contrast, unprotected and can thus decay.  If they decay too early, they may affect the dark matter abundance during or after the time of freeze-out~\cite{Bandyopadhyay:2011qm,Farina:2015uea,Dror:2016rxc,Okawa:2016wrr,Kopp:2016yji}. On the other hand, if they decay too late, they may affect Big Bang Nucleosynthesis (BBN) or the Cosmic Microwave Background (CMB).  In what follows we show that such constraints can easily be satisfied: (I)~Early decays can be avoided for lifetimes longer than the freeze-out time scale of $t_f \approx 10^{-3}$--$10^{-2}$~sec; (II)~Late time decays do not pose a problem since the heavier mesons annihilate efficiently into pions before they decay, and their  Boltzmann-suppressed abundances at the time of decay do not affect BBN or CMB.

\section{Lifetimes}

We now address the decay rates of the unstable heavy twin mesons, $\eta$ and $\theta$. We begin with the $\eta$ meson lifetime. Its decays proceed into two off-shell twin photons via the anomaly diagram, with each twin photon decaying into $e^+e^-$ pairs via the kinetic mixing with the SM photon, or via two loops into a pair of muons (due to helicity suppression). 

\begin{figure}[t]
  \centering
  \begin{fmffile}{triangle}
    \begin{fmfgraph*}(200,60) 
      \fmfleft{i1} \fmfright{o1,o2,o3,o4}
      \fmfforce{(0.1w,0.5h)}{i1}
      \fmfforce{(0.3w,0.5h)}{v1}
      \fmfforce{(0.5w,0.1h)}{v2}
      \fmfforce{(0.5w,0.9h)}{v3}
      \fmfforce{(0.7w,0.1h)}{v4}
      \fmfforce{(0.7w,0.9h)}{v5}
      \fmfforce{(0.6w,0.1h)}{v6}
      \fmfforce{(0.6w,0.9h)}{v7}
      \fmfforce{(0.9w,1h)}{o1}
      \fmfforce{(0.9w,0.8h)}{o2}
      \fmfforce{(0.9w,0.2h)}{o3}
      \fmfforce{(0.9w,0h)}{o4}
      \fmflabel{}{i1}
      \fmflabel{$\eta$}{i1}
      \fmflabel{$e^-$}{o1}
      \fmflabel{$e^+$}{o2}
      \fmflabel{$e^-$}{o3}
      \fmflabel{$e^+$}{o4}
      \fmf{dashes,tension=3}{i1,v1} 
      \fmf{fermion}{v1,v2}
      \fmf{fermion,label=$q'$}{v2,v3}
      \fmf{fermion}{v3,v1}
      \fmf{photon,label=$\gamma'$,l.side=right}{v2,v6}
      \fmf{photon,label=$\gamma'$,l.side=left}{v3,v7}
      \fmf{photon,label=$\gamma^{\phantom{\prime}}$,l.side=right}{v6,v4}
      \fmf{photon,label=$\gamma^{\phantom{\prime}}$,l.side=left}{v7,v5}
      \fmf{fermion}{o2,v5,o1}
      \fmf{fermion}{o4,v4,o3}
      \fmfv{decor.shape=cross,decor.size=0.1h}{v6}
      \fmfv{decor.shape=cross,decor.size=0.1h}{v7}
    \end{fmfgraph*}
  \end{fmffile}
  
  \vspace{4mm}
    \vspace{6mm}
  
%
  \centering
  \begin{fmffile}{box}
    \begin{fmfgraph*}(200,60) 
      \fmfleft{i1} \fmfright{o1,o2}
      \fmfforce{(0.1w,0.5h)}{i1}
      \fmfforce{(0.3w,0.5h)}{v1}
      \fmfforce{(0.5w,0.1h)}{v2}
      \fmfforce{(0.5w,0.9h)}{v3}
      \fmfforce{(0.7w,0.1h)}{v4}
      \fmfforce{(0.7w,0.9h)}{v5}
      \fmfforce{(0.6w,0.1h)}{v6}
      \fmfforce{(0.6w,0.9h)}{v7}
      \fmfforce{(0.9w,0.1h)}{o1}
      \fmfforce{(0.9w,0.9h)}{o2}
      \fmflabel{$\eta$}{i1}
      \fmflabel{$\mu^+$}{o1}
      \fmflabel{$\mu^-$}{o2}
      \fmf{dashes,tension=3}{i1,v1} 
      \fmf{fermion}{v1,v2}
      \fmf{fermion,label=$q'$}{v2,v3}
      \fmf{fermion}{v3,v1}
      \fmf{photon,label=$\gamma'$,l.side=right}{v2,v6}
      \fmf{photon,label=$\gamma'$,l.side=left}{v3,v7}
      \fmf{photon,label=$\gamma^{\phantom{\prime}}$,l.side=right}{v6,v4}
      \fmf{photon,label=$\gamma^{\phantom{\prime}}$,l.side=left}{v7,v5}
      \fmf{fermion}{o2,v5}
      \fmf{fermion}{v5,v4}
      \fmf{fermion}{v4,o1}
      \fmfv{decor.shape=cross,decor.size=0.1h}{v6}
      \fmfv{decor.shape=cross,decor.size=0.1h}{v7}
    \end{fmfgraph*}
  \end{fmffile}
  
  \vspace{4mm}
    \vspace{6mm}
    
  \centering
  \begin{fmffile}{W}
    \begin{fmfgraph*}(200,60) 
      \fmfleft{i1} \fmfright{o1,o2}
      \fmfforce{(0.1w,0.5h)}{i1}
      \fmfforce{(0.2w,0.5h)}{v1}
      \fmfforce{(0.4w,0.1h)}{v2}
      \fmfforce{(0.4w,0.9h)}{v3}
      \fmfforce{(0.6w,0.1h)}{v4}
      \fmfforce{(0.6w,0.9h)}{v5}
      \fmfforce{(0.8w,0.5h)}{v6}
      \fmfforce{(0.7w,0.9h)}{v7}
      \fmfforce{(0.7w,0.1h)}{v8}
      \fmfforce{(0.8w,0.9h)}{v9}
      \fmfforce{(0.8w,0.1h)}{v10}
      \fmfforce{(0.9w,1h)}{o1}
      \fmfforce{(0.9w,0.8h)}{o2}
      \fmfforce{(0.9w,0.2h)}{o3}
      \fmfforce{(0.9w,0h)}{o4}
      \fmfforce{(0.9w,0.5h)}{o5}
      \fmflabel{$\theta$}{i1}
      \fmflabel{$e^-$}{o1}
      \fmflabel{$e^+$}{o2}
      \fmflabel{$e^-$}{o3}
      \fmflabel{$e^+$}{o4}
      \fmflabel{$\pi$}{o5}
      \fmf{dashes,tension=3}{o5,v6} 
      \fmf{dashes,tension=3}{i1,v1} 
      \fmf{fermion,label=$u'\mbox{ or }c'$,l.side=right}{v1,v2}
      \fmf{photon,label=$W$}{v2,v3}
      \fmf{fermion,label=$u'\mbox{ or }c'$,l.side=right}{v3,v1}
      \fmf{fermion,label=$d'\mbox{ or }s'$,l.side=right}{v2,v4}
      \fmf{fermion,label=$d'\mbox{ or }s'$,l.side=right}{v5,v3}
      \fmf{fermion}{v4,v6,v5}
      \fmf{fermion}{o2,v9,o1}
      \fmf{fermion}{o4,v10,o3}
      \fmf{photon,label=$W$}{v2,v3}
      \fmf{photon,label=$\gamma'$,l.side=right}{v4,v8}
      \fmf{photon,label=$\gamma'$,l.side=left}{v5,v7}
      \fmf{photon,label=$\gamma^{\phantom{\prime}}$,l.side=right}{v8,v10}
      \fmf{photon,label=$\gamma^{\phantom{\prime}}$,l.side=left}{v7,v9}
      \fmfv{decor.shape=cross,decor.size=0.1h}{v7}
      \fmfv{decor.shape=cross,decor.size=0.1h}{v8}
    \end{fmfgraph*}
  \end{fmffile}
  
    \vspace{1cm}

  \caption{\label{fig:W}\label{fig:triangle} \label{fig:box} Sample Feynman diagrams for (top to bottom) ${\eta \rightarrow e^+ e^-e^+ e^-}$, $\eta \rightarrow \mu^+ \mu^-$, and $\theta \rightarrow \pi\ e^+ e^-e^+ e^-$.  The cross in the diagrams refers to the kinetic mixing between the twin photon $\gamma'$ and the SM photon $\gamma$.}
\end{figure}
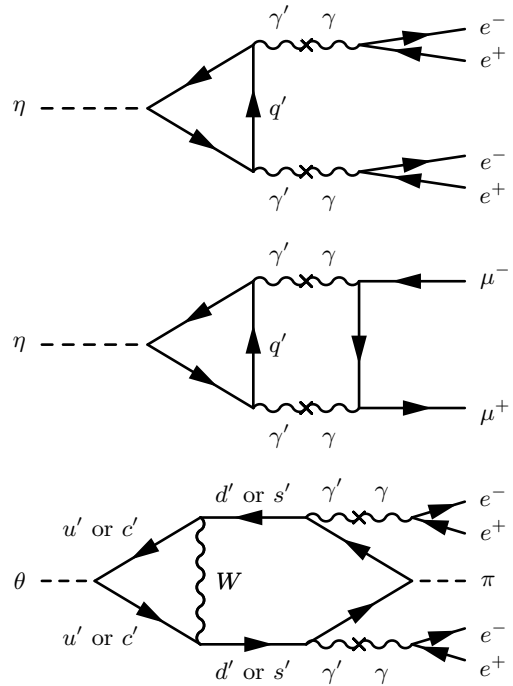

The decay rate of the $\eta$ via the one-loop 4-body process (see top diagram of Fig.~\ref{fig:triangle}) can be estimated by starting with the standard formula for pion decay via the anomaly diagram in the SM, replacing the photons with off-shell dark photons, and further decaying each of those into an $e^+e^-$ pair, which suppresses the decay rate by an additional factor of $\left[\frac{\alpha}{2\pi}\epsilon^2\left(\frac{m_\eta}{2m_{\gamma'}}\right)^4\right]^2.$
We arrive at the estimate
\beq\label{eq:4body}
\lefteqn{\Gamma(\eta\to e^+ e^- e^+ e^-)} \no \\
&\approx &\frac{9\,\alpha_D^2\, m_\eta^3}{8\pi^3 f_\pi^2} \left[\left(\frac{2}{3}\right)^2-\left(\frac{-1}{3}\right)^2\right]^2 \left[\frac{\alpha}{2\pi}\epsilon^2\left(\frac{m_\eta}{2m_{\gamma'}}\right)^4 \right]^2\,\no\\
&=&\frac{1}{ 4 \times 10^{11}\;{\rm sec}  }\left(\frac{\epsilon}{10^{-4}}\right)^4\left(\frac{3\;{\rm GeV}}{m_\gamma'}\right)^8\no\\
&&\quad\quad\times\left(\frac{m_\eta}{300\;{\rm MeV}}\right)^{9}\left(\frac{\alpha_D}{1/4\pi}\right)^2\left(\frac{m_\eta/f_\pi}{2\pi}\right)^2\,.
\eeq
Note that our definition of the decay constant $f_\pi$ follows that in Refs.~\cite{Witten:1983tw,Witten:1983tx,Hochberg:2014kqa} and differs from the notation often used in the SM by a factor of $2\sqrt{2}$.  The decay is fastest for the largest value of kinetic mixing and smallest twin photon mass, $\epsilon \approx 10^{-3}$ and $m'_\gamma \approx 2 m_\pi$.

The decay rate of the $\eta$ via the two-loop 2-body process where the dark photons are closed into a loop  (see middle diagram of Fig.~\ref{fig:box}) is helicity suppressed, and hence the final state will be $\mu^+\mu^-$, with an estimated decay rate
\beq\label{eq:2body}
\lefteqn{
\Gamma(\eta \to \mu^+\mu^-)} \no\\
&\approx &\frac{9\,\alpha_D^2\, m_\eta^3}{8\pi^3 f_\pi^2} \left[\left(\frac{2}{3}\right)^2-\left(\frac{-1}{3}\right)^2\right]^2 \left(\frac{\alpha}{2\pi}\epsilon^2\frac{m_\mu}{m_\pi}\frac{m_\eta^2}{m_{\gamma'}^2}\right)^2\no\\
&= &  \frac{1}{10^{6}\;{\rm sec}}  \left(\frac{\epsilon}{10^{-4}}\right)^4\left(\frac{3\;{\rm GeV}}{m_\gamma'}\right)^4\no\\
&&\quad\quad\times\left(\frac{m_\eta}{300\;{\rm MeV}}\right)^{3}\left(\frac{\alpha_D}{1/4\pi}\right)^2\left(\frac{m_\eta/f_\pi}{2\pi}\right)^2\left(\frac{m_\eta}{m_\pi}\right)^2.
\eeq
While more important for higher $m_{\gamma'}$, this is moderately suppressed compared to Eq.~\eqref{eq:4body} for small $m_{\gamma'}$.

The $\theta$ meson has the same quantum number as $\pi$ under the exact $SU(2)_f$, and hence decays as $\theta \rightarrow \pi \gamma^* \gamma^* \rightarrow \pi +2(e^+ e^-)$ via $t$-channel $W$-exchange  (see bottom diagram of Fig.~\ref{fig:W}) .  We estimate
\beq\label{eq:theta}
\lefteqn{
\Gamma\left(\theta \rightarrow \pi\;e^+ e^-e^+ e^-\right)}\no\\
&\approx & \frac{\alpha_D^2}{8\pi} G_F^{\prime 2} f_\pi^4 m_{\theta} \left[\frac{\alpha}{2\pi}\epsilon^2 \left(\frac{\Delta m_\pi}{2m_{\gamma'}}\right)^4\right]^2\no\\
&= & \frac{1}{2 \times 10^{35}\;{\rm sec}}
\left(\frac{\epsilon}{10^{-4}}\right)^4\left(\frac{\alpha_D}{1/4\pi}\right)^2\left(\frac{\Delta}{10\%}\right)^8 \left(\frac{v}{f}\right)^4 \no\\
&&\quad\quad \times \left(\frac{3\;{\rm GeV}}{m_{\gamma'}}\right)^8\left(\frac{2\pi}{m_\pi/f_\pi}\right)^{4} \left(\frac{m_\pi}{300\;{\rm MeV}}\right)^{13}\,,
\eeq
which is much longer than the age of the Universe.  If $\Delta < 4 m_e/m_\pi$, the decay is kinematically forbidden. For lifetime longer than $10^{27}$~sec, the decay does not lead to an excessive $\gamma$-ray signal from the galactic halo~\cite{Essig:2013goa}.

We learn that the $\eta$ and $\theta$ twin mesons can both be present at the time of freeze-out and participate in the $3\rightarrow 2$ annihilation process. 

\begin{figure}[t!]
\begin{center}
\includegraphics[width =0.46\textwidth]{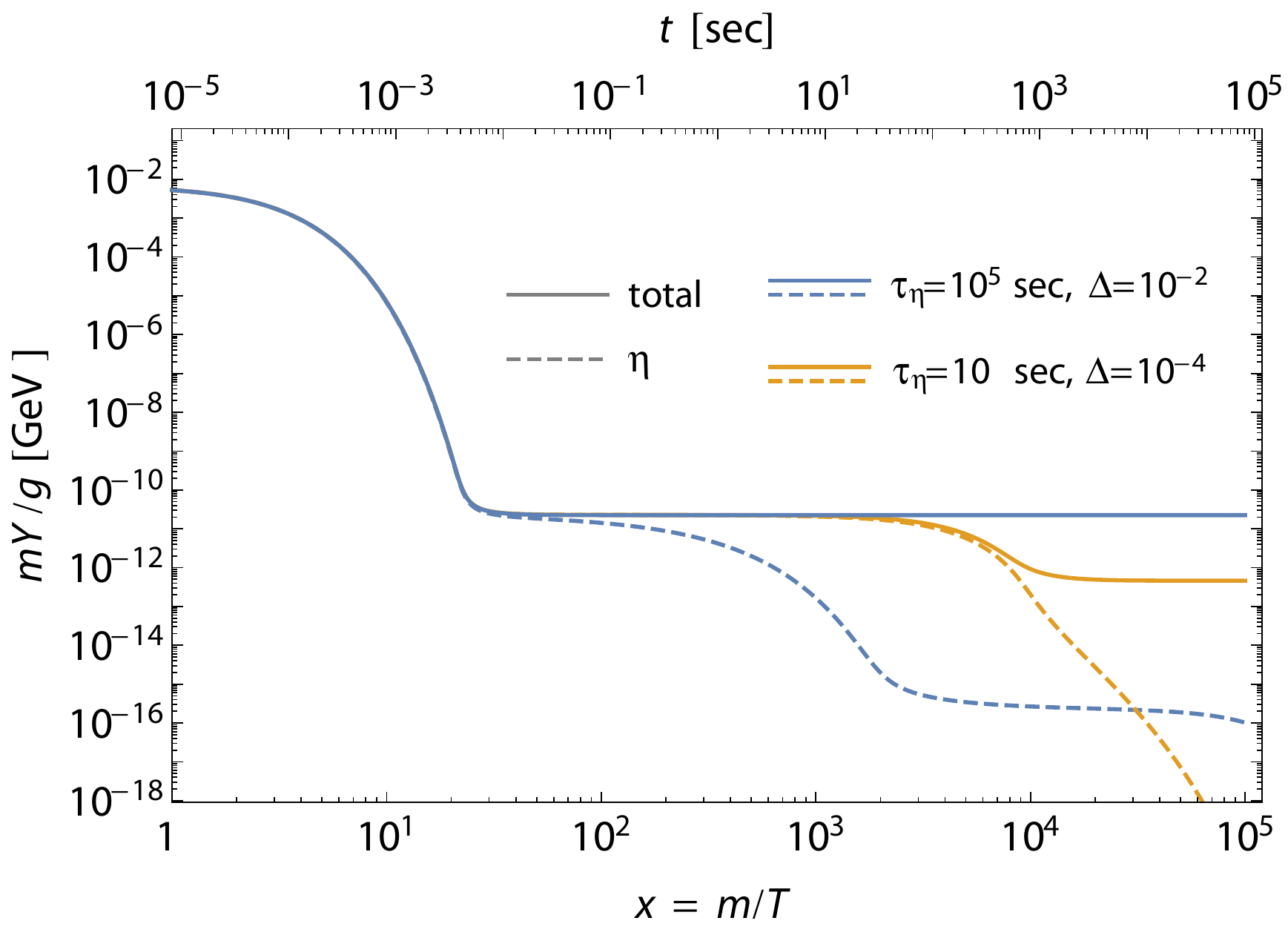}
\end{center}
\caption{\label{fig:schematic}
A schematic description of the sequence of events for twin SIMPs.  Solid (dashed) curves depict the mass times total meson ($\eta$ meson) yield per degree of freedom. At very high temperatures, the $3\rightarrow 2$ annihilation process reduces the abundance of all twin mesons.  At $x=m_\pi/T\simeq 20$, the $3\rightarrow 2$ annihilations freeze out. Two different lifetimes are shown to illustrate distinct scenarios: (1) Blue curves, $\tau_\eta = 10^5~{\rm sec},~ \Delta = 10^{-2}$. The decay happens much after chemical decoupling, in which case the pion abundance is not affected by the decays, and the correct dark matter abundance is set by the $3\to2$ annihilations. (2) Orange curves, $\tau_\eta = 10~{\rm sec},~ \Delta = 10^{-4}$. The $\eta$ decays out of equilibrium with the SM, but decays in chemical equilibrium with the other mesons. In this case the dark matter abundance is depleted via the decays, and then freezes out, leading to too small a relic abundance.}
\end{figure}

After freezeout, the strong interactions among the twin mesons can maintain chemical equilibrium between $\pi$ and heavier mesons.  For instance, with the strong interaction of 
\beq
\langle\sigma v\rangle_{\eta\eta \rightarrow \pi\pi} = \frac{m_\pi^2}{128\pi f_\pi^4} \beta_f,
\quad \beta_f =\sqrt{1-\frac{m_\pi^2}{m_\eta^2}},
\eeq
this process decouples at
\begin{equation}
\langle \sigma v\rangle_{\eta\eta \rightarrow \pi\pi}Y_\pi e^{-2(m_\eta-m_\pi)/T_{\rm chem}} s(T_{\rm chem}) \simeq H(T_{\rm chem}),
\end{equation}
and similarly for other mesons.  We find that the heavier meson abundances are Boltzmann suppressed and do not contribute significantly to the current dark matter density. This can be seen in Fig.~\ref{fig:schematic}, where we present solutions to the Boltzmann equations. Solid curves represent the total mesons, while the dashed curves represent the $\eta$ mesons, for different mass splittings and lifetimes. As is evident, for sufficiently large $\eta$ lifetimes, the pion abundance set by the $3\to2$ annihilations is unaffected by the decays of the heavy $\eta$ meson.  

The decays of $\eta$ can also potentially cause dissociation of light elements after BBN and/or possibly affect the CMB.  If the lifetime of $\eta$ is longer than $10^6$~sec, its decay may cause distortions in the CMB spectrum, depending on how Boltzmann-suppressed the $\eta$ abundance is~\cite{Kawasaki:2017bqm}.  The BBN constraints may be stronger down to $\tau_\eta \sim 10^4$~sec for decays to electrons.  However, the limits have been derived in the literature only for heavy decaying particles, with masses above a few 10's of GeV (see {\it e.g.} Ref.~\cite{Kawasaki:2017bqm}), as opposed to the case at hand where the decaying particles are ${\cal O}(300\ {\rm MeV})$. As a result, the constraints in the literature are not directly applicable here: the actual BBN constraint in our setup is expected to be weaker. We are not aware of detailed constraints on electromagnetic energy injection during the BBN era when the decay products have sub-GeV energies~\cite{private}.  
When taking into account the Boltzmann suppression of the heavy mesons at the time of decay, we have checked that the $\eta$-lifetime estimated in Eqs.~\eqref{eq:4body} and ~\eqref{eq:2body} can easily satisfy the currently derived limits for mass splittings $\Delta \gsim 10^{-4}$, and we expect the accurate allowed range to be even broader as explained above. 

We note that kinetic equilibrium between the twin mesons and SM sectors has been assumed thus far throughout. However, depending on $(m_{\gamma'}, \epsilon)$, the mesons may drop out of kinetic equilibrium with the SM soon after freeze-out occurs. In this case, the meson system cools more rapidly than the SM sector---$T_{\rm meson} \propto a^{-2}$ vs. $T_{\rm SM}\propto a^{-1}$---resulting in a more severe Boltzmann suppression of the heavy meson abundances and weaker constraints.

A lower bound on the mass splitting $\Delta$ arises also from indirect detection data. 
This is because the annihilation process $\pi \pi \rightarrow \eta \eta \rightarrow {\rm SM}$ could, in principle, produce large gamma ray fluxes in the galactic center today.  The current experimental limit is $\langle\sigma v\rangle \lsim10^{-25} {\rm cm}^3 {\rm sec}^{-1}$~\cite{Essig:2013goa}, compared to the naive expectation in our setup of $\langle\sigma v\rangle \approx 10^{-25} {\rm cm}^2 \times v_0 \approx 10^{-17} {\rm cm}^3 {\rm sec}^{-1}$, which would be unacceptably large.  Here, $v_0\sim 220~\mbox{km/sec}$ is the velocity dispersion at the galactic center.  This present-day annihilation process is, however, kinematically forbidden if the mass splitting among the mesons is larger than the typical kinetic energy in the galactic halo, $\Delta \gtrsim v_0^2 \approx 10^{-6}$. 

We conclude that the dark matter is not depleted after freeze-out despite the fact that some of the twin mesons decay into the visible sector. Constraints from BBN and indirect detection are evaded for meson mass splittings $\Delta \gsim 10^{-4}$, while 
for $\Delta \sim 10^{-4}-10^{-6}$, a detailed analysis of the impact on BBN is required for this light dark matter mass range---an effort which is underway~\cite{private}.   It would be interesting to study whether slight modifications to the light element abundances are possible and/or if observable $\gamma$-ray signals can arise due to the Maxwellian tail in the kinematic distribution of mesons in the halo. This and other mechanisms to set the twin pion abundance in this setup beyond the SIMP will be explored in detail in future work~\cite{future}. 

\section{Experimental Signatures}

The strongest limit to date on twin Higgs theories arises from the invisible Higgs width.  In the ${\mathbb Z}_2$-limit, electroweak symmetry breaking in both sectors would occur at the same energy scale $f=v$, and the observed 125~GeV Higgs would contain equal components in both sectors, decaying invisibly into the twin sector 50\% of the time.  Clearly this is not the case; the invisible width of the Higgs dictates $f/v \gtrsim 3$ at present.  The LHC will further improve this limit, and the ILC would probe the invisible width down to the 0.3\% level \cite{Barklow:2017suo}.  We note, however, that the Higgs decay into the twin sector is not entirely invisible in our model.  If the Higgs decays into twin gluons $g'g'$, the $g'$ fragments into mesons.  As discussed above, some of the mesons decay visibly $\theta \rightarrow \pi \;e^+e^-e^+e^-$ and $\eta \rightarrow e^+e^-e^+e^-, \mu^+ \mu^-$, and therefore part of the twin jet is visible with narrowly collimated low-mass lepton pairs.  Some of the twin vector mesons $\rho$ would also mix with the twin photon and lead to $\rho\rightarrow \gamma^{\prime *} \rightarrow \gamma^* \rightarrow l^+ l^-$ and be visible.  Such phenomenology is reminiscent of Hidden Valley models~\cite{Strassler:2006im}.

The twin photon is in essence the kinetically mixed dark photon which has been studied extensively in recent literature.  In the near future, the Belle-II experiment will improve the sensitivity by at least an order of magnitude, substantially cutting into the parameter space~\cite{Soffer:2014ona,Essig:2013vha}.  Belle-II may even do better because their calorimeters do not point exactly to the collision point and they hence can avoid cracks that allow a real photon to escape detection \cite{private3}. Spectroscopy of the resonance states in the twin QCD can be performed as well, using single-photon events at low energy lepton colliders~\cite{Hochberg:2015vrg,Hochberg:2017khi}. This is particularly interesting if the $b'$ is lighter than the $b$ to allow for twin bottomonium spectroscopy.

Twin SIMP dark matter shows promise in direct detection efforts as well.  For example, $D^\pm$ mesons can interact with the SM particles via the exchange of the massive twin photon.  The details depend on the twin meson splitting $\Delta$ that determines the Boltzmann suppression of their abundance.  Potential prospects in this mass range include the use of either electron recoils in semiconductors~\cite{Essig:2011nj,Graham:2012su,Tiffenberg:2017aac,Crisler:2018gci}, atomic ionization~\cite{Essig:2011nj,Essig:2012yx}, superconductors~\cite{Hochberg:2015pha,Hochberg:2015fth}, scintillators~\cite{Derenzo:2016fse} and two-dimensional targets such as graphene~\cite{Hochberg:2016ntt}, or the use of nuclear recoils in color centers~\cite{Essig:2016crl,Budnik:2017sbu} and in superfluid helium~\cite{Schutz:2016tid,danfuture}.

\section{Summary}

We have proposed a model that addresses both the hierarchy problem (via the twin Higgs mechanism) and the dark matter problem (as a Strongly Interacting Massive Particle, or SIMP).  The twin pion-like bound states are the lightest particles in the twin sector, and are stable due an exact global $SU(2)_f$ symmetry between the first two generations.  The relic abundance of dark matter is set through $3\to2$ self-annihilations of the twin mesons via the Wess-Zumino-Witten term. The near coincidence of strong scales between the scale of QCD and that of the SIMP sector, as required in the SIMP mechanism, is naturally explained in such a framework. 

The framework predicts rich and diverse experimental probes into its parameter space.  Dark matter self-scatterings near their current astrophysical limit are expected, which may address small-scale structure puzzles. Precise measurements of the velocity profile of stars in dwarf galaxies using the Prime Focus Spectrograph on the Subaru telescope~\cite{Ellis:2012rn} will further test this.
At high energy colliders, final states with missing energy and soft collimated leptons can be produced, as well as entirely visible final states. Moreover, low-energy $e^+ e^-$ colliders such as Super KEK-B and direct detection searches for dark matter are promising experimental avenues for testing the theory. We leave a detailed study of the phenomenology of this setup to future work \cite{future}.

\mysections{Acknowledgments}
We thank Nathaniel Craig, Torben Ferber, Masahiro Kawasaki, Kazunori Kohri, Dan McKinsey, Takeo Moroi, Maxim Pospelov, Matt Pyle   and Vetri Velan for useful conversations. We also thank Robert McGehee  and Katelin  Schutz for comments on the manuscript. 
 HM thanks the Alexander von Humboldt Foundation for support while this work was completed. HM also thanks the hospitality of the DESY Theory Group despite his unanticipated interruptions during his stay.  The work of YH is supported by the Israel Science Foundation (grant No. 1112/17), by the Binational Science Foundation (grant No. 2016155), by the I-CORE Program of the Planning Budgeting Committee (grant No. 1937/12), by the German Israel Foundation (grant No. I-2487-303.7/2017), and  by the Azrieli Foundation. EK is supported by the Israel Science Foundation (grant No. 1111/17), by the Binational Science Foundation  (grant No. 2016153) and by the I-CORE Program of the Planning Budgeting Committee (grant No. 1937/12).  HM was supported by the NSF under grant PHY-1638509, and by the U.S. DOE under Contract DE-AC02-05CH11231.  HM was also supported by the JSPS Grant-in-Aid for Scientific Research (C) (No.~26400241 and 17K05409), MEXT Grant-in-Aid for Scientific Research on Innovative Areas (No.~15H05887, 15K21733),  by WPI, MEXT, Japan, and  by the Binational Science Foundation  (grant No. 2016153).

\bibliography{bibliospectroscopy}{}

\begin{thebibliography}{50}%
\makeatletter
\providecommand \@ifxundefined [1]{%
 \@ifx{#1\undefined}
}%
\providecommand \@ifnum [1]{%
 \ifnum #1\expandafter \@firstoftwo
 \else \expandafter \@secondoftwo
 \fi
}%
\providecommand \@ifx [1]{%
 \ifx #1\expandafter \@firstoftwo
 \else \expandafter \@secondoftwo
 \fi
}%
\providecommand \natexlab [1]{#1}%
\providecommand \enquote  [1]{``#1''}%
\providecommand \bibnamefont  [1]{#1}%
\providecommand \bibfnamefont [1]{#1}%
\providecommand \citenamefont [1]{#1}%
\providecommand \href@noop [0]{\@secondoftwo}%
\providecommand \href [0]{\begingroup \@sanitize@url \@href}%
\providecommand \@href[1]{\@@startlink{#1}\@@href}%
\providecommand \@@href[1]{\endgroup#1\@@endlink}%
\providecommand \@sanitize@url [0]{\catcode `\\12\catcode `\$12\catcode
  `\&12\catcode `\#12\catcode `\^12\catcode `\_12\catcode `\%12\relax}%
\providecommand \@@startlink[1]{}%
\providecommand \@@endlink[0]{}%
\providecommand \url  [0]{\begingroup\@sanitize@url \@url }%
\providecommand \@url [1]{\endgroup\@href {#1}{\urlprefix }}%
\providecommand \urlprefix  [0]{URL }%
\providecommand \Eprint [0]{\href }%
\providecommand \doibase [0]{http://dx.doi.org/}%
\providecommand \selectlanguage [0]{\@gobble}%
\providecommand \bibinfo  [0]{\@secondoftwo}%
\providecommand \bibfield  [0]{\@secondoftwo}%
\providecommand \translation [1]{[#1]}%
\providecommand \BibitemOpen [0]{}%
\providecommand \bibitemStop [0]{}%
\providecommand \bibitemNoStop [0]{.\EOS\space}%
\providecommand \EOS [0]{\spacefactor3000\relax}%
\providecommand \BibitemShut  [1]{\csname bibitem#1\endcsname}%
\let\auto@bib@innerbib\@empty
\bibitem [{\citenamefont {Chacko}\ \emph
  {et~al.}(2006{\natexlab{a}})\citenamefont {Chacko}, \citenamefont {Goh},\
  and\ \citenamefont {Harnik}}]{Chacko:2005pe}%
  \BibitemOpen
  \bibfield  {author} {\bibinfo {author} {\bibfnamefont {Z.}~\bibnamefont
  {Chacko}}, \bibinfo {author} {\bibfnamefont {H.-S.}\ \bibnamefont {Goh}}, \
  and\ \bibinfo {author} {\bibfnamefont {R.}~\bibnamefont {Harnik}},\ }\href
  {\doibase 10.1103/PhysRevLett.96.231802} {\bibfield  {journal} {\bibinfo
  {journal} {Phys. Rev. Lett.}\ }\textbf {\bibinfo {volume} {96}},\ \bibinfo
  {pages} {231802} (\bibinfo {year} {2006}{\natexlab{a}})},\ \Eprint
  {http://arxiv.org/abs/hep-ph/0506256} {arXiv:hep-ph/0506256 [hep-ph]}
  \BibitemShut {NoStop}%
\bibitem [{\citenamefont {Chacko}\ \emph
  {et~al.}(2006{\natexlab{b}})\citenamefont {Chacko}, \citenamefont {Nomura},
  \citenamefont {Papucci},\ and\ \citenamefont {Perez}}]{Chacko:2005vw}%
  \BibitemOpen
  \bibfield  {author} {\bibinfo {author} {\bibfnamefont {Z.}~\bibnamefont
  {Chacko}}, \bibinfo {author} {\bibfnamefont {Y.}~\bibnamefont {Nomura}},
  \bibinfo {author} {\bibfnamefont {M.}~\bibnamefont {Papucci}}, \ and\
  \bibinfo {author} {\bibfnamefont {G.}~\bibnamefont {Perez}},\ }\href
  {\doibase 10.1088/1126-6708/2006/01/126} {\bibfield  {journal} {\bibinfo
  {journal} {JHEP}\ }\textbf {\bibinfo {volume} {01}},\ \bibinfo {pages} {126}
  (\bibinfo {year} {2006}{\natexlab{b}})},\ \Eprint
  {http://arxiv.org/abs/hep-ph/0510273} {arXiv:hep-ph/0510273 [hep-ph]}
  \BibitemShut {NoStop}%
\bibitem [{\citenamefont {Batell}\ and\ \citenamefont
  {McCullough}(2015)}]{Batell:2015aha}%
  \BibitemOpen
  \bibfield  {author} {\bibinfo {author} {\bibfnamefont {B.}~\bibnamefont
  {Batell}}\ and\ \bibinfo {author} {\bibfnamefont {M.}~\bibnamefont
  {McCullough}},\ }\href {\doibase 10.1103/PhysRevD.92.073018} {\bibfield
  {journal} {\bibinfo  {journal} {Phys. Rev.}\ }\textbf {\bibinfo {volume}
  {D92}},\ \bibinfo {pages} {073018} (\bibinfo {year} {2015})},\ \Eprint
  {http://arxiv.org/abs/1504.04016} {arXiv:1504.04016 [hep-ph]} \BibitemShut
  {NoStop}%
\bibitem [{\citenamefont {Burdman}\ \emph {et~al.}(2007)\citenamefont
  {Burdman}, \citenamefont {Chacko}, \citenamefont {Goh},\ and\ \citenamefont
  {Harnik}}]{Burdman:2006tz}%
  \BibitemOpen
  \bibfield  {author} {\bibinfo {author} {\bibfnamefont {G.}~\bibnamefont
  {Burdman}}, \bibinfo {author} {\bibfnamefont {Z.}~\bibnamefont {Chacko}},
  \bibinfo {author} {\bibfnamefont {H.-S.}\ \bibnamefont {Goh}}, \ and\
  \bibinfo {author} {\bibfnamefont {R.}~\bibnamefont {Harnik}},\ }\href
  {\doibase 10.1088/1126-6708/2007/02/009} {\bibfield  {journal} {\bibinfo
  {journal} {JHEP}\ }\textbf {\bibinfo {volume} {02}},\ \bibinfo {pages} {009}
  (\bibinfo {year} {2007})},\ \Eprint {http://arxiv.org/abs/hep-ph/0609152}
  {arXiv:hep-ph/0609152 [hep-ph]} \BibitemShut {NoStop}%
\bibitem [{\citenamefont {Cai}\ \emph {et~al.}(2009)\citenamefont {Cai},
  \citenamefont {Cheng},\ and\ \citenamefont {Terning}}]{Cai:2008au}%
  \BibitemOpen
  \bibfield  {author} {\bibinfo {author} {\bibfnamefont {H.}~\bibnamefont
  {Cai}}, \bibinfo {author} {\bibfnamefont {H.-C.}\ \bibnamefont {Cheng}}, \
  and\ \bibinfo {author} {\bibfnamefont {J.}~\bibnamefont {Terning}},\ }\href
  {\doibase 10.1088/1126-6708/2009/05/045} {\bibfield  {journal} {\bibinfo
  {journal} {JHEP}\ }\textbf {\bibinfo {volume} {05}},\ \bibinfo {pages} {045}
  (\bibinfo {year} {2009})},\ \Eprint {http://arxiv.org/abs/0812.0843}
  {arXiv:0812.0843 [hep-ph]} \BibitemShut {NoStop}%
\bibitem [{\citenamefont {Craig}\ \emph
  {et~al.}(2015{\natexlab{a}})\citenamefont {Craig}, \citenamefont {Knapen},\
  and\ \citenamefont {Longhi}}]{Craig:2014aea}%
  \BibitemOpen
  \bibfield  {author} {\bibinfo {author} {\bibfnamefont {N.}~\bibnamefont
  {Craig}}, \bibinfo {author} {\bibfnamefont {S.}~\bibnamefont {Knapen}}, \
  and\ \bibinfo {author} {\bibfnamefont {P.}~\bibnamefont {Longhi}},\ }\href
  {\doibase 10.1103/PhysRevLett.114.061803} {\bibfield  {journal} {\bibinfo
  {journal} {Phys. Rev. Lett.}\ }\textbf {\bibinfo {volume} {114}},\ \bibinfo
  {pages} {061803} (\bibinfo {year} {2015}{\natexlab{a}})},\ \Eprint
  {http://arxiv.org/abs/1410.6808} {arXiv:1410.6808 [hep-ph]} \BibitemShut
  {NoStop}%
\bibitem [{\citenamefont {Craig}\ \emph
  {et~al.}(2015{\natexlab{b}})\citenamefont {Craig}, \citenamefont {Knapen},\
  and\ \citenamefont {Longhi}}]{Craig:2014roa}%
  \BibitemOpen
  \bibfield  {author} {\bibinfo {author} {\bibfnamefont {N.}~\bibnamefont
  {Craig}}, \bibinfo {author} {\bibfnamefont {S.}~\bibnamefont {Knapen}}, \
  and\ \bibinfo {author} {\bibfnamefont {P.}~\bibnamefont {Longhi}},\ }\href
  {\doibase 10.1007/JHEP03(2015)106} {\bibfield  {journal} {\bibinfo  {journal}
  {JHEP}\ }\textbf {\bibinfo {volume} {03}},\ \bibinfo {pages} {106} (\bibinfo
  {year} {2015}{\natexlab{b}})},\ \Eprint {http://arxiv.org/abs/1411.7393}
  {arXiv:1411.7393 [hep-ph]} \BibitemShut {NoStop}%
\bibitem [{\citenamefont {Arkani-Hamed}\ \emph {et~al.}(2016)\citenamefont
  {Arkani-Hamed}, \citenamefont {Cohen}, \citenamefont {D'Agnolo},
  \citenamefont {Hook}, \citenamefont {Kim},\ and\ \citenamefont
  {Pinner}}]{Arkani-Hamed:2016rle}%
  \BibitemOpen
  \bibfield  {author} {\bibinfo {author} {\bibfnamefont {N.}~\bibnamefont
  {Arkani-Hamed}}, \bibinfo {author} {\bibfnamefont {T.}~\bibnamefont {Cohen}},
  \bibinfo {author} {\bibfnamefont {R.~T.}\ \bibnamefont {D'Agnolo}}, \bibinfo
  {author} {\bibfnamefont {A.}~\bibnamefont {Hook}}, \bibinfo {author}
  {\bibfnamefont {H.~D.}\ \bibnamefont {Kim}}, \ and\ \bibinfo {author}
  {\bibfnamefont {D.}~\bibnamefont {Pinner}},\ }\href {\doibase
  10.1103/PhysRevLett.117.251801} {\bibfield  {journal} {\bibinfo  {journal}
  {Phys. Rev. Lett.}\ }\textbf {\bibinfo {volume} {117}},\ \bibinfo {pages}
  {251801} (\bibinfo {year} {2016})},\ \Eprint
  {http://arxiv.org/abs/1607.06821} {arXiv:1607.06821 [hep-ph]} \BibitemShut
  {NoStop}%
\bibitem [{\citenamefont {Hochberg}\ \emph {et~al.}(2014)\citenamefont
  {Hochberg}, \citenamefont {Kuflik}, \citenamefont {Volansky},\ and\
  \citenamefont {Wacker}}]{Hochberg:2014dra}%
  \BibitemOpen
  \bibfield  {author} {\bibinfo {author} {\bibfnamefont {Y.}~\bibnamefont
  {Hochberg}}, \bibinfo {author} {\bibfnamefont {E.}~\bibnamefont {Kuflik}},
  \bibinfo {author} {\bibfnamefont {T.}~\bibnamefont {Volansky}}, \ and\
  \bibinfo {author} {\bibfnamefont {J.~G.}\ \bibnamefont {Wacker}},\ }\href
  {\doibase 10.1103/PhysRevLett.113.171301} {\bibfield  {journal} {\bibinfo
  {journal} {Phys. Rev. Lett.}\ }\textbf {\bibinfo {volume} {113}},\ \bibinfo
  {pages} {171301} (\bibinfo {year} {2014})},\ \Eprint
  {http://arxiv.org/abs/1402.5143} {arXiv:1402.5143 [hep-ph]} \BibitemShut
  {NoStop}%
\bibitem [{\citenamefont {Hochberg}\ \emph {et~al.}(2015)\citenamefont
  {Hochberg}, \citenamefont {Kuflik}, \citenamefont {Murayama}, \citenamefont
  {Volansky},\ and\ \citenamefont {Wacker}}]{Hochberg:2014kqa}%
  \BibitemOpen
  \bibfield  {author} {\bibinfo {author} {\bibfnamefont {Y.}~\bibnamefont
  {Hochberg}}, \bibinfo {author} {\bibfnamefont {E.}~\bibnamefont {Kuflik}},
  \bibinfo {author} {\bibfnamefont {H.}~\bibnamefont {Murayama}}, \bibinfo
  {author} {\bibfnamefont {T.}~\bibnamefont {Volansky}}, \ and\ \bibinfo
  {author} {\bibfnamefont {J.~G.}\ \bibnamefont {Wacker}},\ }\href {\doibase
  10.1103/PhysRevLett.115.021301} {\bibfield  {journal} {\bibinfo  {journal}
  {Phys. Rev. Lett.}\ }\textbf {\bibinfo {volume} {115}},\ \bibinfo {pages}
  {021301} (\bibinfo {year} {2015})},\ \Eprint {http://arxiv.org/abs/1411.3727}
  {arXiv:1411.3727 [hep-ph]} \BibitemShut {NoStop}%
\bibitem [{\citenamefont {Wess}\ and\ \citenamefont
  {Zumino}(1971)}]{Wess:1971yu}%
  \BibitemOpen
  \bibfield  {author} {\bibinfo {author} {\bibfnamefont {J.}~\bibnamefont
  {Wess}}\ and\ \bibinfo {author} {\bibfnamefont {B.}~\bibnamefont {Zumino}},\
  }\href {\doibase 10.1016/0370-2693(71)90582-X} {\bibfield  {journal}
  {\bibinfo  {journal} {Phys. Lett.}\ }\textbf {\bibinfo {volume} {B37}},\
  \bibinfo {pages} {95} (\bibinfo {year} {1971})}\BibitemShut {NoStop}%
\bibitem [{\citenamefont {Witten}(1983{\natexlab{a}})}]{Witten:1983tw}%
  \BibitemOpen
  \bibfield  {author} {\bibinfo {author} {\bibfnamefont {E.}~\bibnamefont
  {Witten}},\ }\href {\doibase 10.1016/0550-3213(83)90063-9} {\bibfield
  {journal} {\bibinfo  {journal} {Nucl. Phys.}\ }\textbf {\bibinfo {volume}
  {B223}},\ \bibinfo {pages} {422} (\bibinfo {year}
  {1983}{\natexlab{a}})}\BibitemShut {NoStop}%
\bibitem [{\citenamefont {Witten}(1983{\natexlab{b}})}]{Witten:1983tx}%
  \BibitemOpen
  \bibfield  {author} {\bibinfo {author} {\bibfnamefont {E.}~\bibnamefont
  {Witten}},\ }\href {\doibase 10.1016/0550-3213(83)90064-0} {\bibfield
  {journal} {\bibinfo  {journal} {Nucl. Phys.}\ }\textbf {\bibinfo {volume}
  {B223}},\ \bibinfo {pages} {433} (\bibinfo {year}
  {1983}{\natexlab{b}})}\BibitemShut {NoStop}%
\bibitem [{\citenamefont {Lee}\ and\ \citenamefont {Seo}(2015)}]{Lee:2015gsa}%
  \BibitemOpen
  \bibfield  {author} {\bibinfo {author} {\bibfnamefont {H.~M.}\ \bibnamefont
  {Lee}}\ and\ \bibinfo {author} {\bibfnamefont {M.-S.}\ \bibnamefont {Seo}},\
  }\href {\doibase 10.1016/j.physletb.2015.07.013} {\bibfield  {journal}
  {\bibinfo  {journal} {Phys. Lett.}\ }\textbf {\bibinfo {volume} {B748}},\
  \bibinfo {pages} {316} (\bibinfo {year} {2015})},\ \Eprint
  {http://arxiv.org/abs/1504.00745} {arXiv:1504.00745 [hep-ph]} \BibitemShut
  {NoStop}%
\bibitem [{\citenamefont {Hochberg}\ \emph
  {et~al.}(2016{\natexlab{a}})\citenamefont {Hochberg}, \citenamefont
  {Kuflik},\ and\ \citenamefont {Murayama}}]{Hochberg:2015vrg}%
  \BibitemOpen
  \bibfield  {author} {\bibinfo {author} {\bibfnamefont {Y.}~\bibnamefont
  {Hochberg}}, \bibinfo {author} {\bibfnamefont {E.}~\bibnamefont {Kuflik}}, \
  and\ \bibinfo {author} {\bibfnamefont {H.}~\bibnamefont {Murayama}},\ }\href
  {\doibase 10.1007/JHEP05(2016)090} {\bibfield  {journal} {\bibinfo  {journal}
  {JHEP}\ }\textbf {\bibinfo {volume} {05}},\ \bibinfo {pages} {090} (\bibinfo
  {year} {2016}{\natexlab{a}})},\ \Eprint {http://arxiv.org/abs/1512.07917}
  {arXiv:1512.07917 [hep-ph]} \BibitemShut {NoStop}%
\bibitem [{\citenamefont {Berlin}\ \emph {et~al.}(2018)\citenamefont {Berlin},
  \citenamefont {Blinov}, \citenamefont {Gori}, \citenamefont {Schuster},\ and\
  \citenamefont {Toro}}]{Berlin:2018tvf}%
  \BibitemOpen
  \bibfield  {author} {\bibinfo {author} {\bibfnamefont {A.}~\bibnamefont
  {Berlin}}, \bibinfo {author} {\bibfnamefont {N.}~\bibnamefont {Blinov}},
  \bibinfo {author} {\bibfnamefont {S.}~\bibnamefont {Gori}}, \bibinfo {author}
  {\bibfnamefont {P.}~\bibnamefont {Schuster}}, \ and\ \bibinfo {author}
  {\bibfnamefont {N.}~\bibnamefont {Toro}},\ }\href@noop {} {\  (\bibinfo
  {year} {2018})},\ \Eprint {http://arxiv.org/abs/1801.05805} {arXiv:1801.05805
  [hep-ph]} \BibitemShut {NoStop}%
\bibitem [{\citenamefont {García~García}\ \emph
  {et~al.}(2015{\natexlab{a}})\citenamefont {García~García}, \citenamefont
  {Lasenby},\ and\ \citenamefont {March-Russell}}]{Garcia:2015loa}%
  \BibitemOpen
  \bibfield  {author} {\bibinfo {author} {\bibfnamefont {I.}~\bibnamefont
  {García~García}}, \bibinfo {author} {\bibfnamefont {R.}~\bibnamefont
  {Lasenby}}, \ and\ \bibinfo {author} {\bibfnamefont {J.}~\bibnamefont
  {March-Russell}},\ }\href {\doibase 10.1103/PhysRevD.92.055034} {\bibfield
  {journal} {\bibinfo  {journal} {Phys. Rev.}\ }\textbf {\bibinfo {volume}
  {D92}},\ \bibinfo {pages} {055034} (\bibinfo {year} {2015}{\natexlab{a}})},\
  \Eprint {http://arxiv.org/abs/1505.07109} {arXiv:1505.07109 [hep-ph]}
  \BibitemShut {NoStop}%
\bibitem [{\citenamefont {Craig}\ and\ \citenamefont
  {Katz}(2015)}]{Craig:2015xla}%
  \BibitemOpen
  \bibfield  {author} {\bibinfo {author} {\bibfnamefont {N.}~\bibnamefont
  {Craig}}\ and\ \bibinfo {author} {\bibfnamefont {A.}~\bibnamefont {Katz}},\
  }\href {\doibase 10.1088/1475-7516/2015/10/054} {\bibfield  {journal}
  {\bibinfo  {journal} {JCAP}\ }\textbf {\bibinfo {volume} {1510}},\ \bibinfo
  {pages} {054} (\bibinfo {year} {2015})},\ \Eprint
  {http://arxiv.org/abs/1505.07113} {arXiv:1505.07113 [hep-ph]} \BibitemShut
  {NoStop}%
\bibitem [{\citenamefont {García~García}\ \emph
  {et~al.}(2015{\natexlab{b}})\citenamefont {García~García}, \citenamefont
  {Lasenby},\ and\ \citenamefont {March-Russell}}]{Garcia:2015toa}%
  \BibitemOpen
  \bibfield  {author} {\bibinfo {author} {\bibfnamefont {I.}~\bibnamefont
  {García~García}}, \bibinfo {author} {\bibfnamefont {R.}~\bibnamefont
  {Lasenby}}, \ and\ \bibinfo {author} {\bibfnamefont {J.}~\bibnamefont
  {March-Russell}},\ }\href {\doibase 10.1103/PhysRevLett.115.121801}
  {\bibfield  {journal} {\bibinfo  {journal} {Phys. Rev. Lett.}\ }\textbf
  {\bibinfo {volume} {115}},\ \bibinfo {pages} {121801} (\bibinfo {year}
  {2015}{\natexlab{b}})},\ \Eprint {http://arxiv.org/abs/1505.07410}
  {arXiv:1505.07410 [hep-ph]} \BibitemShut {NoStop}%
\bibitem [{\citenamefont {Farina}(2015)}]{Farina:2015uea}%
  \BibitemOpen
  \bibfield  {author} {\bibinfo {author} {\bibfnamefont {M.}~\bibnamefont
  {Farina}},\ }\href {\doibase 10.1088/1475-7516/2015/11/017} {\bibfield
  {journal} {\bibinfo  {journal} {JCAP}\ }\textbf {\bibinfo {volume} {1511}},\
  \bibinfo {pages} {017} (\bibinfo {year} {2015})},\ \Eprint
  {http://arxiv.org/abs/1506.03520} {arXiv:1506.03520 [hep-ph]} \BibitemShut
  {NoStop}%
\bibitem [{\citenamefont {Craig}\ \emph
  {et~al.}(2015{\natexlab{c}})\citenamefont {Craig}, \citenamefont {Katz},
  \citenamefont {Strassler},\ and\ \citenamefont {Sundrum}}]{Craig:2015pha}%
  \BibitemOpen
  \bibfield  {author} {\bibinfo {author} {\bibfnamefont {N.}~\bibnamefont
  {Craig}}, \bibinfo {author} {\bibfnamefont {A.}~\bibnamefont {Katz}},
  \bibinfo {author} {\bibfnamefont {M.}~\bibnamefont {Strassler}}, \ and\
  \bibinfo {author} {\bibfnamefont {R.}~\bibnamefont {Sundrum}},\ }\href
  {\doibase 10.1007/JHEP07(2015)105} {\bibfield  {journal} {\bibinfo  {journal}
  {JHEP}\ }\textbf {\bibinfo {volume} {07}},\ \bibinfo {pages} {105} (\bibinfo
  {year} {2015}{\natexlab{c}})},\ \Eprint {http://arxiv.org/abs/1501.05310}
  {arXiv:1501.05310 [hep-ph]} \BibitemShut {NoStop}%
\bibitem [{\citenamefont {Geller}\ \emph {et~al.}(2018)\citenamefont {Geller},
  \citenamefont {Iwamoto}, \citenamefont {Lee}, \citenamefont {Shadmi},\ and\
  \citenamefont {Telem}}]{Geller:2018biy}%
  \BibitemOpen
  \bibfield  {author} {\bibinfo {author} {\bibfnamefont {M.}~\bibnamefont
  {Geller}}, \bibinfo {author} {\bibfnamefont {S.}~\bibnamefont {Iwamoto}},
  \bibinfo {author} {\bibfnamefont {G.}~\bibnamefont {Lee}}, \bibinfo {author}
  {\bibfnamefont {Y.}~\bibnamefont {Shadmi}}, \ and\ \bibinfo {author}
  {\bibfnamefont {O.}~\bibnamefont {Telem}},\ }\href@noop {} {\  (\bibinfo
  {year} {2018})},\ \Eprint {http://arxiv.org/abs/1802.07720} {arXiv:1802.07720
  [hep-ph]} \BibitemShut {NoStop}%
\bibitem [{\citenamefont {Kuflik}\ \emph {et~al.}(2015)\citenamefont {Kuflik},
  \citenamefont {Perelstein}, \citenamefont {Lorier},\ and\ \citenamefont
  {Tsai}}]{Kuflik:2015isi}%
  \BibitemOpen
  \bibfield  {author} {\bibinfo {author} {\bibfnamefont {E.}~\bibnamefont
  {Kuflik}}, \bibinfo {author} {\bibfnamefont {M.}~\bibnamefont {Perelstein}},
  \bibinfo {author} {\bibfnamefont {N.~R.-L.}\ \bibnamefont {Lorier}}, \ and\
  \bibinfo {author} {\bibfnamefont {Y.-D.}\ \bibnamefont {Tsai}},\ }\href@noop
  {} {\  (\bibinfo {year} {2015})},\ \Eprint {http://arxiv.org/abs/1512.04545}
  {arXiv:1512.04545 [hep-ph]} \BibitemShut {NoStop}%
\bibitem [{\citenamefont {Kuflik}\ \emph {et~al.}(2017)\citenamefont {Kuflik},
  \citenamefont {Perelstein}, \citenamefont {Lorier},\ and\ \citenamefont
  {Tsai}}]{Kuflik:2017iqs}%
  \BibitemOpen
  \bibfield  {author} {\bibinfo {author} {\bibfnamefont {E.}~\bibnamefont
  {Kuflik}}, \bibinfo {author} {\bibfnamefont {M.}~\bibnamefont {Perelstein}},
  \bibinfo {author} {\bibfnamefont {N.~R.-L.}\ \bibnamefont {Lorier}}, \ and\
  \bibinfo {author} {\bibfnamefont {Y.-D.}\ \bibnamefont {Tsai}},\ }\href
  {\doibase 10.1007/JHEP08(2017)078} {\bibfield  {journal} {\bibinfo  {journal}
  {JHEP}\ }\textbf {\bibinfo {volume} {08}},\ \bibinfo {pages} {078} (\bibinfo
  {year} {2017})},\ \Eprint {http://arxiv.org/abs/1706.05381} {arXiv:1706.05381
  [hep-ph]} \BibitemShut {NoStop}%
\bibitem [{\citenamefont {Bandyopadhyay}\ \emph {et~al.}(2011)\citenamefont
  {Bandyopadhyay}, \citenamefont {Chun},\ and\ \citenamefont
  {Park}}]{Bandyopadhyay:2011qm}%
  \BibitemOpen
  \bibfield  {author} {\bibinfo {author} {\bibfnamefont {P.}~\bibnamefont
  {Bandyopadhyay}}, \bibinfo {author} {\bibfnamefont {E.~J.}\ \bibnamefont
  {Chun}}, \ and\ \bibinfo {author} {\bibfnamefont {J.-C.}\ \bibnamefont
  {Park}},\ }\href {\doibase 10.1007/JHEP06(2011)129} {\bibfield  {journal}
  {\bibinfo  {journal} {JHEP}\ }\textbf {\bibinfo {volume} {06}},\ \bibinfo
  {pages} {129} (\bibinfo {year} {2011})},\ \Eprint
  {http://arxiv.org/abs/1105.1652} {arXiv:1105.1652 [hep-ph]} \BibitemShut
  {NoStop}%
\bibitem [{\citenamefont {Dror}\ \emph {et~al.}(2016)\citenamefont {Dror},
  \citenamefont {Kuflik},\ and\ \citenamefont {Ng}}]{Dror:2016rxc}%
  \BibitemOpen
  \bibfield  {author} {\bibinfo {author} {\bibfnamefont {J.~A.}\ \bibnamefont
  {Dror}}, \bibinfo {author} {\bibfnamefont {E.}~\bibnamefont {Kuflik}}, \ and\
  \bibinfo {author} {\bibfnamefont {W.~H.}\ \bibnamefont {Ng}},\ }\href
  {\doibase 10.1103/PhysRevLett.117.211801} {\bibfield  {journal} {\bibinfo
  {journal} {Phys. Rev. Lett.}\ }\textbf {\bibinfo {volume} {117}},\ \bibinfo
  {pages} {211801} (\bibinfo {year} {2016})},\ \Eprint
  {http://arxiv.org/abs/1607.03110} {arXiv:1607.03110 [hep-ph]} \BibitemShut
  {NoStop}%
\bibitem [{\citenamefont {Okawa}\ \emph {et~al.}(2017)\citenamefont {Okawa},
  \citenamefont {Tanabashi},\ and\ \citenamefont {Yamanaka}}]{Okawa:2016wrr}%
  \BibitemOpen
  \bibfield  {author} {\bibinfo {author} {\bibfnamefont {S.}~\bibnamefont
  {Okawa}}, \bibinfo {author} {\bibfnamefont {M.}~\bibnamefont {Tanabashi}}, \
  and\ \bibinfo {author} {\bibfnamefont {M.}~\bibnamefont {Yamanaka}},\ }\href
  {\doibase 10.1103/PhysRevD.95.023006} {\bibfield  {journal} {\bibinfo
  {journal} {Phys. Rev.}\ }\textbf {\bibinfo {volume} {D95}},\ \bibinfo {pages}
  {023006} (\bibinfo {year} {2017})},\ \Eprint
  {http://arxiv.org/abs/1607.08520} {arXiv:1607.08520 [hep-ph]} \BibitemShut
  {NoStop}%
\bibitem [{\citenamefont {Kopp}\ \emph {et~al.}(2016)\citenamefont {Kopp},
  \citenamefont {Liu}, \citenamefont {Slatyer}, \citenamefont {Wang},\ and\
  \citenamefont {Xue}}]{Kopp:2016yji}%
  \BibitemOpen
  \bibfield  {author} {\bibinfo {author} {\bibfnamefont {J.}~\bibnamefont
  {Kopp}}, \bibinfo {author} {\bibfnamefont {J.}~\bibnamefont {Liu}}, \bibinfo
  {author} {\bibfnamefont {T.~R.}\ \bibnamefont {Slatyer}}, \bibinfo {author}
  {\bibfnamefont {X.-P.}\ \bibnamefont {Wang}}, \ and\ \bibinfo {author}
  {\bibfnamefont {W.}~\bibnamefont {Xue}},\ }\href {\doibase
  10.1007/JHEP12(2016)033} {\bibfield  {journal} {\bibinfo  {journal} {JHEP}\
  }\textbf {\bibinfo {volume} {12}},\ \bibinfo {pages} {033} (\bibinfo {year}
  {2016})},\ \Eprint {http://arxiv.org/abs/1609.02147} {arXiv:1609.02147
  [hep-ph]} \BibitemShut {NoStop}%
\bibitem [{\citenamefont {Essig}\ \emph
  {et~al.}(2013{\natexlab{a}})\citenamefont {Essig}, \citenamefont {Kuflik},
  \citenamefont {McDermott}, \citenamefont {Volansky},\ and\ \citenamefont
  {Zurek}}]{Essig:2013goa}%
  \BibitemOpen
  \bibfield  {author} {\bibinfo {author} {\bibfnamefont {R.}~\bibnamefont
  {Essig}}, \bibinfo {author} {\bibfnamefont {E.}~\bibnamefont {Kuflik}},
  \bibinfo {author} {\bibfnamefont {S.~D.}\ \bibnamefont {McDermott}}, \bibinfo
  {author} {\bibfnamefont {T.}~\bibnamefont {Volansky}}, \ and\ \bibinfo
  {author} {\bibfnamefont {K.~M.}\ \bibnamefont {Zurek}},\ }\href {\doibase
  10.1007/JHEP11(2013)193} {\bibfield  {journal} {\bibinfo  {journal} {JHEP}\
  }\textbf {\bibinfo {volume} {11}},\ \bibinfo {pages} {193} (\bibinfo {year}
  {2013}{\natexlab{a}})},\ \Eprint {http://arxiv.org/abs/1309.4091}
  {arXiv:1309.4091 [hep-ph]} \BibitemShut {NoStop}%
\bibitem [{\citenamefont {Kawasaki}\ \emph {et~al.}(2018)\citenamefont
  {Kawasaki}, \citenamefont {Kohri}, \citenamefont {Moroi},\ and\ \citenamefont
  {Takaesu}}]{Kawasaki:2017bqm}%
  \BibitemOpen
  \bibfield  {author} {\bibinfo {author} {\bibfnamefont {M.}~\bibnamefont
  {Kawasaki}}, \bibinfo {author} {\bibfnamefont {K.}~\bibnamefont {Kohri}},
  \bibinfo {author} {\bibfnamefont {T.}~\bibnamefont {Moroi}}, \ and\ \bibinfo
  {author} {\bibfnamefont {Y.}~\bibnamefont {Takaesu}},\ }\href {\doibase
  10.1103/PhysRevD.97.023502} {\bibfield  {journal} {\bibinfo  {journal} {Phys.
  Rev.}\ }\textbf {\bibinfo {volume} {D97}},\ \bibinfo {pages} {023502}
  (\bibinfo {year} {2018})},\ \Eprint {http://arxiv.org/abs/1709.01211}
  {arXiv:1709.01211 [hep-ph]} \BibitemShut {NoStop}%
\bibitem [{\citenamefont {Kawasaki}\ \emph {et~al.}(tion)\citenamefont
  {Kawasaki}, \citenamefont {Kohri},\ and\ \citenamefont {Moroi}}]{private}%
  \BibitemOpen
  \bibfield  {author} {\bibinfo {author} {\bibfnamefont {M.}~\bibnamefont
  {Kawasaki}}, \bibinfo {author} {\bibfnamefont {K.}~\bibnamefont {Kohri}}, \
  and\ \bibinfo {author} {\bibfnamefont {T.}~\bibnamefont {Moroi}},\
  }\href@noop {} {\  (\bibinfo {year} {{private communication}})}\BibitemShut
  {NoStop}%
\bibitem [{\citenamefont {Hochberg}\ \emph {et~al.}()\citenamefont {Hochberg},
  \citenamefont {Kuflik}, \citenamefont {McGehee}, \citenamefont {Murayama},\
  and\ \citenamefont {Schutz}}]{future}%
  \BibitemOpen
  \bibfield  {author} {\bibinfo {author} {\bibfnamefont {Y.}~\bibnamefont
  {Hochberg}}, \bibinfo {author} {\bibfnamefont {E.}~\bibnamefont {Kuflik}},
  \bibinfo {author} {\bibfnamefont {R.}~\bibnamefont {McGehee}}, \bibinfo
  {author} {\bibfnamefont {H.}~\bibnamefont {Murayama}}, \ and\ \bibinfo
  {author} {\bibfnamefont {K.}~\bibnamefont {Schutz}},\ }\href@noop {} {\
  }\Eprint {http://arxiv.org/abs/work in progress} {work in progress}
  \BibitemShut {NoStop}%
\bibitem [{\citenamefont {Barklow}\ \emph {et~al.}(2018)\citenamefont
  {Barklow}, \citenamefont {Fujii}, \citenamefont {Jung}, \citenamefont {Karl},
  \citenamefont {List}, \citenamefont {Ogawa}, \citenamefont {Peskin},\ and\
  \citenamefont {Tian}}]{Barklow:2017suo}%
  \BibitemOpen
  \bibfield  {author} {\bibinfo {author} {\bibfnamefont {T.}~\bibnamefont
  {Barklow}}, \bibinfo {author} {\bibfnamefont {K.}~\bibnamefont {Fujii}},
  \bibinfo {author} {\bibfnamefont {S.}~\bibnamefont {Jung}}, \bibinfo {author}
  {\bibfnamefont {R.}~\bibnamefont {Karl}}, \bibinfo {author} {\bibfnamefont
  {J.}~\bibnamefont {List}}, \bibinfo {author} {\bibfnamefont {T.}~\bibnamefont
  {Ogawa}}, \bibinfo {author} {\bibfnamefont {M.~E.}\ \bibnamefont {Peskin}}, \
  and\ \bibinfo {author} {\bibfnamefont {J.}~\bibnamefont {Tian}},\ }\href
  {\doibase 10.1103/PhysRevD.97.053003} {\bibfield  {journal} {\bibinfo
  {journal} {Phys. Rev.}\ }\textbf {\bibinfo {volume} {D97}},\ \bibinfo {pages}
  {053003} (\bibinfo {year} {2018})},\ \Eprint
  {http://arxiv.org/abs/1708.08912} {arXiv:1708.08912 [hep-ph]} \BibitemShut
  {NoStop}%
\bibitem [{\citenamefont {Strassler}\ and\ \citenamefont
  {Zurek}(2007)}]{Strassler:2006im}%
  \BibitemOpen
  \bibfield  {author} {\bibinfo {author} {\bibfnamefont {M.~J.}\ \bibnamefont
  {Strassler}}\ and\ \bibinfo {author} {\bibfnamefont {K.~M.}\ \bibnamefont
  {Zurek}},\ }\href {\doibase 10.1016/j.physletb.2007.06.055} {\bibfield
  {journal} {\bibinfo  {journal} {Phys. Lett.}\ }\textbf {\bibinfo {volume}
  {B651}},\ \bibinfo {pages} {374} (\bibinfo {year} {2007})},\ \Eprint
  {http://arxiv.org/abs/hep-ph/0604261} {arXiv:hep-ph/0604261 [hep-ph]}
  \BibitemShut {NoStop}%
\bibitem [{\citenamefont {Soffer}(2014)}]{Soffer:2014ona}%
  \BibitemOpen
  \bibfield  {author} {\bibinfo {author} {\bibfnamefont {A.}~\bibnamefont
  {Soffer}}\ }(\bibinfo {year} {2014})\ \Eprint
  {http://arxiv.org/abs/1409.5263} {arXiv:1409.5263 [hep-ex]} \BibitemShut
  {NoStop}%
\bibitem [{\citenamefont {Essig}\ \emph
  {et~al.}(2013{\natexlab{b}})\citenamefont {Essig}, \citenamefont {Mardon},
  \citenamefont {Papucci}, \citenamefont {Volansky},\ and\ \citenamefont
  {Zhong}}]{Essig:2013vha}%
  \BibitemOpen
  \bibfield  {author} {\bibinfo {author} {\bibfnamefont {R.}~\bibnamefont
  {Essig}}, \bibinfo {author} {\bibfnamefont {J.}~\bibnamefont {Mardon}},
  \bibinfo {author} {\bibfnamefont {M.}~\bibnamefont {Papucci}}, \bibinfo
  {author} {\bibfnamefont {T.}~\bibnamefont {Volansky}}, \ and\ \bibinfo
  {author} {\bibfnamefont {Y.-M.}\ \bibnamefont {Zhong}},\ }\href {\doibase
  10.1007/JHEP11(2013)167} {\bibfield  {journal} {\bibinfo  {journal} {JHEP}\
  }\textbf {\bibinfo {volume} {11}},\ \bibinfo {pages} {167} (\bibinfo {year}
  {2013}{\natexlab{b}})},\ \Eprint {http://arxiv.org/abs/1309.5084}
  {arXiv:1309.5084 [hep-ph]} \BibitemShut {NoStop}%
\bibitem [{\citenamefont {Ferber}(tion)}]{private3}%
  \BibitemOpen
  \bibfield  {author} {\bibinfo {author} {\bibfnamefont {T.}~\bibnamefont
  {Ferber}},\ }\href@noop {} {\  (\bibinfo {year} {{private
  communication}})}\BibitemShut {NoStop}%
\bibitem [{\citenamefont {Hochberg}\ \emph {et~al.}(2018)\citenamefont
  {Hochberg}, \citenamefont {Kuflik},\ and\ \citenamefont
  {Murayama}}]{Hochberg:2017khi}%
  \BibitemOpen
  \bibfield  {author} {\bibinfo {author} {\bibfnamefont {Y.}~\bibnamefont
  {Hochberg}}, \bibinfo {author} {\bibfnamefont {E.}~\bibnamefont {Kuflik}}, \
  and\ \bibinfo {author} {\bibfnamefont {H.}~\bibnamefont {Murayama}},\ }\href
  {\doibase 10.1103/PhysRevD.97.055030} {\bibfield  {journal} {\bibinfo
  {journal} {Phys. Rev.}\ }\textbf {\bibinfo {volume} {D97}},\ \bibinfo {pages}
  {055030} (\bibinfo {year} {2018})},\ \Eprint
  {http://arxiv.org/abs/1706.05008} {arXiv:1706.05008 [hep-ph]} \BibitemShut
  {NoStop}%
\bibitem [{\citenamefont {Essig}\ \emph
  {et~al.}(2012{\natexlab{a}})\citenamefont {Essig}, \citenamefont {Mardon},\
  and\ \citenamefont {Volansky}}]{Essig:2011nj}%
  \BibitemOpen
  \bibfield  {author} {\bibinfo {author} {\bibfnamefont {R.}~\bibnamefont
  {Essig}}, \bibinfo {author} {\bibfnamefont {J.}~\bibnamefont {Mardon}}, \
  and\ \bibinfo {author} {\bibfnamefont {T.}~\bibnamefont {Volansky}},\ }\href
  {\doibase 10.1103/PhysRevD.85.076007} {\bibfield  {journal} {\bibinfo
  {journal} {Phys. Rev.}\ }\textbf {\bibinfo {volume} {D85}},\ \bibinfo {pages}
  {076007} (\bibinfo {year} {2012}{\natexlab{a}})},\ \Eprint
  {http://arxiv.org/abs/1108.5383} {arXiv:1108.5383 [hep-ph]} \BibitemShut
  {NoStop}%
\bibitem [{\citenamefont {Graham}\ \emph {et~al.}(2012)\citenamefont {Graham},
  \citenamefont {Kaplan}, \citenamefont {Rajendran},\ and\ \citenamefont
  {Walters}}]{Graham:2012su}%
  \BibitemOpen
  \bibfield  {author} {\bibinfo {author} {\bibfnamefont {P.~W.}\ \bibnamefont
  {Graham}}, \bibinfo {author} {\bibfnamefont {D.~E.}\ \bibnamefont {Kaplan}},
  \bibinfo {author} {\bibfnamefont {S.}~\bibnamefont {Rajendran}}, \ and\
  \bibinfo {author} {\bibfnamefont {M.~T.}\ \bibnamefont {Walters}},\ }\href
  {\doibase 10.1016/j.dark.2012.09.001} {\bibfield  {journal} {\bibinfo
  {journal} {Phys. Dark Univ.}\ }\textbf {\bibinfo {volume} {1}},\ \bibinfo
  {pages} {32} (\bibinfo {year} {2012})},\ \Eprint
  {http://arxiv.org/abs/1203.2531} {arXiv:1203.2531 [hep-ph]} \BibitemShut
  {NoStop}%
\bibitem [{\citenamefont {Essig}\ \emph
  {et~al.}(2012{\natexlab{b}})\citenamefont {Essig}, \citenamefont
  {Manalaysay}, \citenamefont {Mardon}, \citenamefont {Sorensen},\ and\
  \citenamefont {Volansky}}]{Essig:2012yx}%
  \BibitemOpen
  \bibfield  {author} {\bibinfo {author} {\bibfnamefont {R.}~\bibnamefont
  {Essig}}, \bibinfo {author} {\bibfnamefont {A.}~\bibnamefont {Manalaysay}},
  \bibinfo {author} {\bibfnamefont {J.}~\bibnamefont {Mardon}}, \bibinfo
  {author} {\bibfnamefont {P.}~\bibnamefont {Sorensen}}, \ and\ \bibinfo
  {author} {\bibfnamefont {T.}~\bibnamefont {Volansky}},\ }\href {\doibase
  10.1103/PhysRevLett.109.021301} {\bibfield  {journal} {\bibinfo  {journal}
  {Phys. Rev. Lett.}\ }\textbf {\bibinfo {volume} {109}},\ \bibinfo {pages}
  {021301} (\bibinfo {year} {2012}{\natexlab{b}})},\ \Eprint
  {http://arxiv.org/abs/1206.2644} {arXiv:1206.2644 [astro-ph.CO]} \BibitemShut
  {NoStop}%
\bibitem [{\citenamefont {Hochberg}\ \emph
  {et~al.}(2016{\natexlab{b}})\citenamefont {Hochberg}, \citenamefont {Zhao},\
  and\ \citenamefont {Zurek}}]{Hochberg:2015pha}%
  \BibitemOpen
  \bibfield  {author} {\bibinfo {author} {\bibfnamefont {Y.}~\bibnamefont
  {Hochberg}}, \bibinfo {author} {\bibfnamefont {Y.}~\bibnamefont {Zhao}}, \
  and\ \bibinfo {author} {\bibfnamefont {K.~M.}\ \bibnamefont {Zurek}},\ }\href
  {\doibase 10.1103/PhysRevLett.116.011301} {\bibfield  {journal} {\bibinfo
  {journal} {Phys. Rev. Lett.}\ }\textbf {\bibinfo {volume} {116}},\ \bibinfo
  {pages} {011301} (\bibinfo {year} {2016}{\natexlab{b}})},\ \Eprint
  {http://arxiv.org/abs/1504.07237} {arXiv:1504.07237 [hep-ph]} \BibitemShut
  {NoStop}%
\bibitem [{\citenamefont {Hochberg}\ \emph
  {et~al.}(2016{\natexlab{c}})\citenamefont {Hochberg}, \citenamefont {Pyle},
  \citenamefont {Zhao},\ and\ \citenamefont {Zurek}}]{Hochberg:2015fth}%
  \BibitemOpen
  \bibfield  {author} {\bibinfo {author} {\bibfnamefont {Y.}~\bibnamefont
  {Hochberg}}, \bibinfo {author} {\bibfnamefont {M.}~\bibnamefont {Pyle}},
  \bibinfo {author} {\bibfnamefont {Y.}~\bibnamefont {Zhao}}, \ and\ \bibinfo
  {author} {\bibfnamefont {K.~M.}\ \bibnamefont {Zurek}},\ }\href {\doibase
  10.1007/JHEP08(2016)057} {\bibfield  {journal} {\bibinfo  {journal} {JHEP}\
  }\textbf {\bibinfo {volume} {08}},\ \bibinfo {pages} {057} (\bibinfo {year}
  {2016}{\natexlab{c}})},\ \Eprint {http://arxiv.org/abs/1512.04533}
  {arXiv:1512.04533 [hep-ph]} \BibitemShut {NoStop}%
\bibitem [{\citenamefont {Derenzo}\ \emph {et~al.}(2017)\citenamefont
  {Derenzo}, \citenamefont {Essig}, \citenamefont {Massari}, \citenamefont
  {Soto},\ and\ \citenamefont {Yu}}]{Derenzo:2016fse}%
  \BibitemOpen
  \bibfield  {author} {\bibinfo {author} {\bibfnamefont {S.}~\bibnamefont
  {Derenzo}}, \bibinfo {author} {\bibfnamefont {R.}~\bibnamefont {Essig}},
  \bibinfo {author} {\bibfnamefont {A.}~\bibnamefont {Massari}}, \bibinfo
  {author} {\bibfnamefont {A.}~\bibnamefont {Soto}}, \ and\ \bibinfo {author}
  {\bibfnamefont {T.-T.}\ \bibnamefont {Yu}},\ }\href {\doibase
  10.1103/PhysRevD.96.016026} {\bibfield  {journal} {\bibinfo  {journal} {Phys.
  Rev.}\ }\textbf {\bibinfo {volume} {D96}},\ \bibinfo {pages} {016026}
  (\bibinfo {year} {2017})},\ \Eprint {http://arxiv.org/abs/1607.01009}
  {arXiv:1607.01009 [hep-ph]} \BibitemShut {NoStop}%
\bibitem [{\citenamefont {Hochberg}\ \emph {et~al.}(2017)\citenamefont
  {Hochberg}, \citenamefont {Kahn}, \citenamefont {Lisanti}, \citenamefont
  {Tully},\ and\ \citenamefont {Zurek}}]{Hochberg:2016ntt}%
  \BibitemOpen
  \bibfield  {author} {\bibinfo {author} {\bibfnamefont {Y.}~\bibnamefont
  {Hochberg}}, \bibinfo {author} {\bibfnamefont {Y.}~\bibnamefont {Kahn}},
  \bibinfo {author} {\bibfnamefont {M.}~\bibnamefont {Lisanti}}, \bibinfo
  {author} {\bibfnamefont {C.~G.}\ \bibnamefont {Tully}}, \ and\ \bibinfo
  {author} {\bibfnamefont {K.~M.}\ \bibnamefont {Zurek}},\ }\href {\doibase
  10.1016/j.physletb.2017.06.051} {\bibfield  {journal} {\bibinfo  {journal}
  {Phys. Lett.}\ }\textbf {\bibinfo {volume} {B772}},\ \bibinfo {pages} {239}
  (\bibinfo {year} {2017})},\ \Eprint {http://arxiv.org/abs/1606.08849}
  {arXiv:1606.08849 [hep-ph]} \BibitemShut {NoStop}%
\bibitem [{\citenamefont {Essig}\ \emph {et~al.}(2017)\citenamefont {Essig},
  \citenamefont {Mardon}, \citenamefont {Slone},\ and\ \citenamefont
  {Volansky}}]{Essig:2016crl}%
  \BibitemOpen
  \bibfield  {author} {\bibinfo {author} {\bibfnamefont {R.}~\bibnamefont
  {Essig}}, \bibinfo {author} {\bibfnamefont {J.}~\bibnamefont {Mardon}},
  \bibinfo {author} {\bibfnamefont {O.}~\bibnamefont {Slone}}, \ and\ \bibinfo
  {author} {\bibfnamefont {T.}~\bibnamefont {Volansky}},\ }\href {\doibase
  10.1103/PhysRevD.95.056011} {\bibfield  {journal} {\bibinfo  {journal} {Phys.
  Rev.}\ }\textbf {\bibinfo {volume} {D95}},\ \bibinfo {pages} {056011}
  (\bibinfo {year} {2017})},\ \Eprint {http://arxiv.org/abs/1608.02940}
  {arXiv:1608.02940 [hep-ph]} \BibitemShut {NoStop}%
\bibitem [{\citenamefont {Budnik}\ \emph {et~al.}(2017)\citenamefont {Budnik},
  \citenamefont {Chesnovsky}, \citenamefont {Slone},\ and\ \citenamefont
  {Volansky}}]{Budnik:2017sbu}%
  \BibitemOpen
  \bibfield  {author} {\bibinfo {author} {\bibfnamefont {R.}~\bibnamefont
  {Budnik}}, \bibinfo {author} {\bibfnamefont {O.}~\bibnamefont {Chesnovsky}},
  \bibinfo {author} {\bibfnamefont {O.}~\bibnamefont {Slone}}, \ and\ \bibinfo
  {author} {\bibfnamefont {T.}~\bibnamefont {Volansky}},\ }\href@noop {} {\
  (\bibinfo {year} {2017})},\ \Eprint {http://arxiv.org/abs/1705.03016}
  {arXiv:1705.03016 [hep-ph]} \BibitemShut {NoStop}%
\bibitem [{\citenamefont {Schutz}\ and\ \citenamefont
  {Zurek}(2016)}]{Schutz:2016tid}%
  \BibitemOpen
  \bibfield  {author} {\bibinfo {author} {\bibfnamefont {K.}~\bibnamefont
  {Schutz}}\ and\ \bibinfo {author} {\bibfnamefont {K.~M.}\ \bibnamefont
  {Zurek}},\ }\href {\doibase 10.1103/PhysRevLett.117.121302} {\bibfield
  {journal} {\bibinfo  {journal} {Phys. Rev. Lett.}\ }\textbf {\bibinfo
  {volume} {117}},\ \bibinfo {pages} {121302} (\bibinfo {year} {2016})},\
  \Eprint {http://arxiv.org/abs/1604.08206} {arXiv:1604.08206 [hep-ph]}
  \BibitemShut {NoStop}%
\bibitem [{\citenamefont {Hertel}\ \emph {et~al.}()\citenamefont {Hertel},
  \citenamefont {Biekert}, \citenamefont {Lin}, \citenamefont {Velan},\ and\
  \citenamefont {McKinsey}}]{danfuture}%
  \BibitemOpen
  \bibfield  {author} {\bibinfo {author} {\bibfnamefont {S.}~\bibnamefont
  {Hertel}}, \bibinfo {author} {\bibfnamefont {A.}~\bibnamefont {Biekert}},
  \bibinfo {author} {\bibfnamefont {J.}~\bibnamefont {Lin}}, \bibinfo {author}
  {\bibfnamefont {V.}~\bibnamefont {Velan}}, \ and\ \bibinfo {author}
  {\bibfnamefont {D.~N.}\ \bibnamefont {McKinsey}},\ }\href@noop {} {\ }\Eprint
  {http://arxiv.org/abs/to appear} {to appear} \BibitemShut {NoStop}%
\bibitem [{\citenamefont {Ellis}\ \emph {et~al.}(2014)\citenamefont {Ellis}
  \emph {et~al.}}]{Ellis:2012rn}%
  \BibitemOpen
  \bibfield  {author} {\bibinfo {author} {\bibfnamefont {R.}~\bibnamefont
  {Ellis}} \emph {et~al.} (\bibinfo {collaboration} {PFS Team}),\ }\href
  {\doibase 10.1093/pasj/pst019} {\bibfield  {journal} {\bibinfo  {journal}
  {Publ. Astron. Soc. Jap.}\ }\textbf {\bibinfo {volume} {66}},\ \bibinfo
  {pages} {R1} (\bibinfo {year} {2014})},\ \Eprint
  {http://arxiv.org/abs/1206.0737} {arXiv:1206.0737 [astro-ph.CO]} \BibitemShut
  {NoStop}%
\end{thebibliography}%

\end{document}